\documentclass[11pt,fleqn]{article}

\usepackage{amssymb,amsmath}
\usepackage{natbib}
\usepackage{vmargin,setspace,verbatim}
\usepackage{graphicx,graphics,color,tikz,tikz-network}
\usepackage{microtype}

\usepackage{libertine}
\usepackage[libertine]{newtxmath}

\definecolor{darkgreen}{rgb}{0,0.2,0}
\definecolor{darkred}{rgb}{0.3,0,0}

\usepackage[colorlinks=true, pdfstartview=FitV, linkcolor=darkgreen, citecolor=darkred, urlcolor=black]{hyperref}

\newcounter{llst}
\newenvironment{abet}{\begin{list}{\rm (\alph{llst})}{\usecounter{llst}
\setlength{\itemindent}{0em} \setlength{\leftmargin}{3em}
\setlength{\labelwidth}{2em} \setlength{\labelsep}{1em}}}{\end{list}}
\newenvironment{numm}{\begin{list}{\rm (\roman{llst})}{\usecounter{llst}
\setlength{\itemindent}{0em} \setlength{\leftmargin}{3.5em}
\setlength{\labelwidth}{2.5em} \setlength{\labelsep}{1em}}}{\end{list}}

\newtheorem{theorem}{Theorem}[section]

\newtheorem{corollary}[theorem]{Corollary}

\newtheorem{definition}[theorem]{Definition}
\newtheorem{expl}[theorem]{Example}

\newtheorem{lemma}[theorem]{Lemma}

\newtheorem{proposition}[theorem]{Proposition}

\newtheorem{dscrpt}[theorem]{Description}

\newcounter{axiomatiser}

\newcounter{myclaimcount}

\newenvironment{proof}[1][Proof]{\noindent \textbf{#1.} }{\hfill
\rule{0.5em}{0.5em}}

\newenvironment{example}{\begin{expl} \rm}{\hfill $\blacklozenge$
\end{expl}}

\setpapersize{A4}
\setmarginsrb{80pt}{40pt}{80pt}{60pt}{12pt}{10pt}{12pt}{30pt}

\begin{document}

%\begin{titlepage}

\title{\textbf{Gately Values of Cooperative Games}\thanks{We thank Jean-Jacques Herings, Herv\'e Moulin as well as participants of SING17, the 2023 Workshop on Games \& Networks at QUB, Belfast, and EWET 2023 participants for helpful comments on previous drafts of this paper. Financial support by University of Naples Federico II  through Research Grant  FRA2022--GOAPT  is gratefully acknowledged.}}

\author{Robert P.~Gilles\thanks{Department of Economics, The Queen's University of Belfast, Riddel Hall, 185 Stranmillis Road, Belfast, BT9~5EE, UK. \textsf{Email: r.gilles@qub.ac.uk}} \and Lina Mallozzi\thanks{Department of Mathematics and Applications, University of Naples Federico II, Via Claudio 21, 80125 Naples, Italy. E-mail: \textsf{mallozzi@unina.it}} }

\date{\textsf{In memory of Stef Tijs (1937 -- 2023)} \\[2ex] July 2023}

\maketitle

\begin{abstract}
\singlespace
\noindent
We investigate Gately's solution concept for cooperative games with transferable utilities. Gately's conception introduced a bargaining solution that minimises the maximal quantified ``propensity to disrupt'' the negotiation process of the players over the allocation of the generated collective payoffs. Gately's solution concept is well-defined for a broad class of games. We also consider a generalisation based on a parameter-based quantification of the propensity to disrupt. Furthermore, we investigate the relationship of these generalised Gately values with the Core and the Nucleolus and show that Gately's solution is in the Core for all regular 3-player games. We identify exact conditions under which generally these Gately values are Core imputations for arbitrary regular cooperative games. Finally, we investigate the relationship of the Gately value with the Shapley value. 
\end{abstract}

\begin{description}
\singlespace
\item[Keywords:] Game Theory; cooperative game; sharing value; Gately point; Core.

\item[JEL classification:] C71
\end{description}

\thispagestyle{empty}

\newpage

\setcounter{page}{1} \pagenumbering{arabic}

\section{Introduction: Gately's solution concept}

\citet{Gately1974} seminally considered an allocation method founded on individual players' opportunities to disrupt the negotiations regarding the allocation of the generated collective payoffs. This conception is akin to the underlying logic of the Core \citep{Gillies1959} of a cooperative game: Coalitions of players would threaten to abandon the negotiations over the allocation of the total generated collective worth, if the offered allocation does not at least assign each coalition's worth in the cooperative game.\footnote{It is, therefore, implicitly assumed that a coalition can generate its worth independently of the actions taken by players and coalitions outside that coalition.} Gately's conception formalises \emph{to what extend} coalitions would disrupt such negotiations.

Gately introduced the notion of an individual player's \emph{propensity to disrupt}, expressing the relative disruption an individual player causes when leaving the negotiations. In fact, Gately formulated this ``propensity to disrupt'' as the ratio of the other players' collective loss and the individual player's loss due to disruption of the negotiations. The prevailing solution method aims to minimise the maximal propensity to disrupt over all imputations and players in the game. \citet{Gately2019} show that for most cooperative games this solution method results in a unique imputation, which we can denote as the \emph{Gately value} of the game under consideration.

Clearly, Gately's solution concept falls within the category of a \emph{bargaining-based} solution concepts \citep{Maschler1992} that also encompasses, e.g., the bargaining set \citep{AumannMaschler1964}, the Kernel \citep{DavisMascher1965}, and the Nucleolus \citep{Schmeidler1969}.\footnote{For an overview of these solution concepts we also refer to textbooks such as \citet{Moulin1986,Moulin2004}, \citet{Maschler2013}, \citet{OwenBook} and \citet{Gilles2010}.} Contrary to many of these bargaining-based solution concepts, Gately's conception results in an easily to compute allocation rule that can also be categorised as a \emph{compromise value} such as the CIS-value \citep{DriessenFunaki1991} and the $\tau$-value \citep{Tijs1981}. These solution concepts have a fundamentally different axiomatic foundation than the \emph{fairness-based} allocation rules such as the egalitarian solution, the Shapley value \citep{Shapley1953}, the Banzhaf value \citep{Banzhaf1965,Lehrer1988}, and related notions.

 \citet{Gately1974} investigated his conception in the setting of one particular 3-player cost game only. Gately's notion was extended to arbitrary $n$-player cooperative games by \citet{Gately1976}. \footnote{Littlechild and Vaidya used their extended notion of propensity to disrupt to define the \emph{Disruption Nucleolus} in which these propensities are lexicographically minimised.} \citet{Gately1978} introduced various concepts that are closely related to an extended notion of the propensity to disrupt. They introduced \emph{mollifiers} and \emph{homomollifiers}, measuring the disparities emerging from abandonment of negotiations as differences rather than ratios. These formulations result in associated games with a given cooperative game. \citet{Gately1978} primarily investigated the properties of these associated games. 

\paragraph{Gately points: Existence, uniqueness and relationship with the Core}

\citet{Gately2019} point out that the original research questions as posed by \citet{Gately1974} were never properly investigated and answered in the literature. In particular, Staudacher and Anwander focussed on one particular application within the broad range of possibilities in Gately's approach, namely the so-called \emph{Gately point}---defined as an imputation in which all propensities to disrupt are balanced and minimal. The Gately point is a solution to a minimax problem and Staudacher and Anwander show that every standard cooperative game has a unique Gately point. This settles indeed the most basic question concerning Gately's original conception.

Here, we provide a natural and constructive characterisation of the Gately value based on three simple properties. These properties relate to the properties introduced by \citet{Tijs1987} to characterise the $\tau$-value for quasi-balanced cooperative games. We modify these properties for the class of regular cooperative games to fully axiomatise the Gately value for this class of games. This shows in detail that the Gately value is actually a compromise value on the class of regular cooperative games. 

Finally, we introduce the \emph{dual Gately value} as the Gately point of the dual of a given cooperative game. We show that the dual Gately value is identical to the Gately value for the broad class of regular games. Hence, the Gately value is self-dual.

Gately's definition of his propensity to disrupt puts equal weight on assessing the loss or gain of the other players versus the loss or gain of the player under consideration. We consider a parametric formulation in which a weight is attached to the relative importance of the gain or loss of the individual player in comparison with the weight attached to the gain or loss of all other players in the game. The higher the assigned weight, the more an individual's loss or gain due to disruption is taken into account. 

The imputations that balance these weighted propensities to disrupt are now referred to as generalised $\alpha$-Gately points, where $\alpha >0$ is the weight put on an individual's loss or gain due to disruption. It is clear that $\alpha =1$ refers to the original Gately point. We show that for all $\alpha >0$, all regular cooperative games admit a unique $\alpha$-Gately point, generalising the insight of \citet{Gately2019}. 

For any $\alpha >0$, the unique $\alpha$-Gately point is in the Core of the game if and only if the game satisfies $\alpha$-Top Dominance, a parametric variant of the top convexity condition. In particular, the $\alpha$-Top Dominance condition implies that the game is regular as well as partitionally superadditive. However, counterexamples show that there exist superadditive games with non-empty Cores that do not contain any $\alpha$-Gately point. 

\paragraph{Analysis of the Gately value and other solution concepts}

The main contribution of this paper is the investigation of the relationship between the Gately value and other solution concepts on the class of regular cooperative games. In particular, we explore the Gately value as a Core selector and under what conditions the Gately and Shapley values coincide.

\citet{Gately1974} states clearly that he views the conception of Gately points and related concepts based on his notion of ``propensity to disrupt'' as leading to Core selectors. This is exemplified by the underlying conception of Gately's solution method as a Core-based bargaining process. In particular, this is supported by the analysis of the cost games considered by \citet{Gately1974}, which have rather large Cores. Here, we primarily explore the interesting and yet unexplored relationship between Gately points and the Core. In particular, we show that the unique Gately point is a Core selector for \emph{every} regular 3-player cooperative game. 

\citet{Gately1976} already showed that this cannot be extended to $n$-player games by constructing a 4-player game in which the Gately point is not in the Core. Nevertheless we are able to establish the exact conditions for which the unique Gately point is in the Core of a regular $n$-player game. We refer to this condition as \emph{top dominance} in which the grand coalition generates the largest net benefit in relation to the marginal player contributions.  This condition reduces to \emph{top convexity} for zero-normalised games \citep{Shubik1982,JacksonNouweland2005}.

The axiomatic solution concept seminally introduced by \citet{Shapley1953} is now widely accepted as the prime value for cooperative games. It has resulted in a vast literature on determining the Shapley value and its properties on certain classes of cooperative games such as communication situations \citep{Myerson1977,Myerson1980}, network games \citep{JacksonWolinsky1996}, and hierarchical permission structures \citep{GillesOwenBrink1992,GillesOwen1999}. This indicates the validity of the question for which subclasses of cooperative games an alternative solution concept is equivalent to the Shapley value.

We investigate this equivalence property for the Gately value and conclude that for certain classes of regular games the Gately value results in exactly the same imputation as the Shapley value. This includes the class of cooperative games generated by unanimity games of equal-sized coalitions, the so-called the subclasses of $k$-games \citep{Brink2023}. Other classes of highly regular games also possess this equivalence property, showing that potentially for many subclasses of highly regular cooperative games these values might coincide. 

In particular, the Gately and Shapley values are equivalent on the specific subclass of 2-games. This equivalence enables us to utilise the characterisation developed in \citet[Theorem 1]{Brink2023} to establish an axiomatisation of the Gately value within this particular relevant subclass of regular games. In fact, the 2-games subset encompasses the depiction of productive interaction on networks, where binary value-generating activity occurs on the links forming the network. The insight that many values are equivalent on this particular subclass of 2-games is relevant in the analysis of these games.

\paragraph{Structure of the paper}

We introduce and illustrate Gately's approach through a simple application to a trade problem with three traders in Section 2. Section 3 develops the formal treatment of Gately's approach, defines generalised Gately points and values, and discusses the dual of the Gately value. This section concludes with a natural and simple axiomatisation of the Gately value. Section 4 is devoted to the investigation of the relationship of Gately points, the Core and the Nucleolus for 3-player games. We conclude the paper in Section 5 with the investigation of the Gately value with the Core for arbitrary $n$-player games and the relationship between the Gately and Shapley values.

\section{An illustrative example: Bargaining of a sale}

To illustrate the ideas behind \citet{Gately1974}'s conception, consider a situation with one seller---denoted as player S---and two buyers---denoted, respectively, as players B1 and B2. The seller owns an indivisible object that has value to all three players. For the seller S, the object has an intrinsic value of 1, introducing this as a reservation value in any bargained trade between the seller S and any buyer. Buyer B1 assigns a value of 3 to the object, while buyer B2 attributes a value of 2 to the object. Hence, the buyers have a ceiling on their bids of 3 and 2, respectively.

The seller S can trade the object to either buyer B1 or buyer B2. The buyer has to agree to pay a price $P \geqslant 0$ for acquiring the object. If seller S and buyer B1 negotiate a settlement price of $P \geqslant 0$, the result is a gain from trade of $P-1$ for seller S and a gain from trade of $3-P$ for buyer B1. Clearly, the only viable trades are made at prices that would result in non-negative gains from trade for both seller S and buyer B1. Therefore, the viable range of trades would be $1 \leqslant P \leqslant 3$. Similarly, a trade between seller S and buyer B2 results in gains from trade of $P-1$ and $2-P$, respectively. Note that the viable range for such a trade is given as $1 \leqslant P \leqslant 2$.

\paragraph{Representation the trade situation as a cooperative game}

We can represent this bargaining situation as a cooperative game $v$ on $\{ S, B1 , B2 \}$.\footnote{We emphasise that this bargaining situation can also be represented as a non-cooperative game, as is pursued through a well-developed and major literature. Here the focus is on a TU-representation in which the main question is what a proper and fair allocation of the generated gains from trade would be.} Indeed, if there is no trade and seller S retains the ownership rights, he retains the intrinsic value assigned to the object. Hence, seller S by herself generates a personal value of $v(S)=1$. Similarly, for both buyers $v(B1) = v(B2) =0$ if there is no trade. 

The total value generated from a trade between seller S and buyer B1 at any viable price $P$ is now the sum of the intrinsic value of the object plus the total gains from trade. Hence, the generated total wealth is given by $v(S,B1) =  1+(P-1) + (3-P) = 3$ irrespective of the agreed terms of trade in the viable range $1 \leqslant P \leqslant 3$. Similarly, a trade between seller S and buyer B2 at a viable price $P$ generates a collective wealth of $v(S,B2) = 1+ (P-1) + (2-P) = 2$ for any $1 \leqslant P \leqslant 2$. 

If seller S negotiates trade with both buyers B1 and B2 simultaneously, there would be no further gains than the gains from trade generated from a trade between seller S and buyer B1. Now, the total value from a trade resulting from the grand negotiation would be $v(S,B1,B2)=v(S,B1)=3$.

Therefore, this representation of this trade situation as a cooperative game is fully described by $v(S) =1$, $v(B1) = v(B2)=0$, $v(S,B1) = 3$, $v(S,B2)=2$, $v(B1,B2)=0$ and $v(S,B1,B2)=3$. 

%\paragraph{Describing feasible trades as imputations}

\medskip\noindent
An \emph{imputation} in this game theoretic representation is a triple $(x_s,x_1,x_2) \in \mathbb R^3_+$ such that $x_s+x_1+x_2 = v(S,B1,B2) =3$ and $x_i \geqslant v(i)$ for $i=s,1,2$, where $x_s \geqslant 1$ represents the payout assigned to seller S, $x_1 \geqslant 0$ the payout made to buyer B1, and $x_2 \geqslant 0$ the payout made to buyer B2, respectively. These payouts are assumed to be negotiated between the three trading parties and, therefore, add up to the total value generated from their interaction.

The \emph{Core} of this game consists of all imputations such that $x_s+x_1 \geqslant v(S,B1) =3$ and $x_s+x_2 \geqslant v(S,B2)=2$. Here the value generated between from the trade between seller S and buyer B2 is used as an outside option for seller S, providing a lower bound of $v(S,B2)=2$ on the allocation to seller S. The Core can therefore be determined as $C(v) = \{ (t,3-t,0) \mid 2 \leqslant t \leqslant 3 \}$. 

We compute the \emph{Shapley value} \citep{Shapley1953} of this  trade situation as $\varphi (v) = \left( \, 2 \tfrac{1}{6} , \tfrac{2}{3} , \tfrac{1}{6} \, \right)$. We remark that the Shapley value is \emph{not} a Core allocation, i.e., $\varphi (v) \notin C(v)$.

Finally, we compute the \emph{Nucleolus} \citep{Schmeidler1969} of this trade situation, which is a well-known selector of the Core. In fact, $\mathcal{N} (v) = \left( 2 \tfrac{1}{2} , \tfrac{1}{2} , 0 \right) \in C (v)$ is actually the central imputation of the Core of this game as it is the middle point of the set $C(v)$ as a line piece in $\mathbb R^3$. 

\paragraph{Gately's conception}

\citet{Gately1974} introduced the idea that during a negotiation between the seller and the two buyers, each of these three parties can disrupt the proceedings by departing the negotiations. Gately's analysis was explicitly introduced to delineate and focus on a Core selector in the 3-player application explored in \citet{Gately1974}.\footnote{From the stated methodology, Gately's focus is on how individual players negotiate on the allocation of collectively generated worth. The methodology is similar to the definition of the Core as those allocations to which there are no objections, in the sense that there are no coalitional incentives to abandon the negotiations to allocate the collectively generated worth by pursuing an alternative arrangement.} Gately pursued a much simpler solution than the Nucleolus, which is notoriously hard to compute \citep{Maschler1992}.

Gately founded his approach on an innovative method to describe the negotiation process of allocating the worth generated in the cooperative situation. His focal point was on the disruption of the negotiation process that each individual player can cause. This disruption can be measured through a player's ``propensity to disrupt''. \citet[p.~200--201]{Gately1974} introduces this concept as ``the ratio of how much the two other players would lose if a player would refuse to cooperate to how much that player would lose if it refused to cooperate''. 

If the negotiators consider a proposed imputation $(x_s,x_1,x_2) \geqslant (1,0,0)$ with $x_s+x_1+x_2=v(N)=3$, Gately's notion of the propensity to disrupt by seller S would then be the ratio of the buyers' potential loss $x_1 + x_2 - v(B1,B2)$ to the seller's potential loss from non-cooperation, computed as $x_s - v(S)$. Hence, using $x_s+x_1+x_2 = v(N) =3$, the seller S's propensity to disrupt is
\[
d_s (x_s,x_1,x_2) = \frac{x_1+x_2-v(B1,B2)}{x_s - v(S)} = \frac{x_1+x_2}{x_s-1} = \frac{3-x_s}{x_s-1} = \frac{2}{x_s-1}-1 .
\]
Similarly, we construct the propensity to disrupt for both buyers as
\begin{align*}
	d_1 (x_s,x_1,x_2) & = \frac{x_s+x_2-v(S,B2)}{x_1-v(B1)} = \frac{x_s+x_2-2}{x_1} = \frac{1}{x_1}-1 \\[1ex]
	d_2 (x_s,x_1,x_2) & = \frac{x_s+x_1-v(S,B1)}{x_2-v(B2)} = \frac{x_s+x_1-3}{x_2} = -1
\end{align*}
Gately's motivation was that, if a player would get a relatively small payout, the player's propensity to disrupt the agreement is relatively high.\footnote{In particular, if a player would not get any benefit in the negations in the sense that $x_i = v(i)$, her propensity to disrupt is infinitely large. Similarly, if the player would be proposed to receive the total generated benefit $x_i = v(N)$, her propensity to disrupt is usually negative.} Now, the stated objective of Gately's proposed solution is to select an imputation that minimises the maximum propensity to disrupt at that imputation. Hence, one should select an imputation that solves the minimax problem stated as
\[
\min_{(x_s,x_1,x_2) \geqslant (1,0,0) \colon x_s+x_1+x_2=3} \ \max \, \left\{ \, d_s (x_s,x_1,x_2) , d_1 (x_s,x_1,x_2) , d_2 (x_s,x_1,x_2) \, \right\} .
\]
This is clearly resulting in the requirement that $d_s (x_s,x_1,x_2) = d_1 (x_s,x_1,x_2)$, since $d_2 (x_s,x_1,x_2) =-1$ is certainly not a maximum. This results in Gately's solution to be determined by the equations
\[
	\frac{2}{x_s-1} = \frac{1}{x_1} \qquad \mbox{and} \qquad x_s+x_1 =3 \qquad \mbox{with } x_2=0.
\]
Therefore, the solution to Gately's minimax problem is unique and determined as $g (v) = (2 \tfrac{1}{3}, \tfrac{2}{3} ,0)$. We remark that $g (v) \in C(v)$ is indeed a Core allocation for this particular example. 

\paragraph{Generalising Gately's solution conception}

We propose a generalisation of Gately's conception in which the definition of a player's propensity to disrupt is modified by imposing a weight on the denominator. This weight represents an intensity parameter that measures how much more or less weight a player puts on her own loss in relation to the loss of all other players. Hence, we assume that a player can discount her own losses due to disruption or, conversely, assign more weight to her own losses than the losses of the other players. 

Formally, we introduce a weight parameter $\beta >0$ for the denominator in Gately's propensity to disrupt. Instead of applying this directly to the formulated propensity to disrupt $d_i$, $i = s,1,2$, itself, we apply this weight in the modified form $\rho_i = d_i +1$. Hence, for each of the three parties in the trading situation considered, we introduce the $\beta$-weighted propensity to disrupt as
\[
\rho^\beta_s (x_s,x_1,x_2) = \frac{2}{(x_s-1)^\beta} >0 \qquad \rho^\beta_1 (x_s,x_1,x_2) = \frac{1}{x_1^\beta} >0 \qquad \rho^\beta_s (x_s,x_1,x_2) = 0
\]
For any $\beta >0$, a generalised Gately solution would solve the minimax problem given by
\[
\min_{(x_s,x_1,x_2) \geqslant (1,0,0) \colon x_s+x_1+x_2=3} \ \max \left\{ \, \rho^\beta_s (x_s,x_1,x_2) , \rho^\beta_1 (x_s,x_1,x_2) , \rho^\beta_2 (x_s,x_1,x_2) \, \right\} .
\]
The generalised solution for this modified Gately conception is now determined by the equations:
\[
	\frac{2}{(x_s-1)^\beta} = \frac{1}{x_1^\beta} \qquad \mbox{and} \qquad x_s+x_1 =3 \qquad \mbox{with } x_2=0.
\]
The equations stated above lead to the conclusion that, for every $\beta >0$, the generalised Gately solution is given as $g^\beta (v) = \left( \frac{1+3 \sqrt[\beta]{2}}{1+\sqrt[\beta]{2}} \, , \, \frac{2}{1+\sqrt[\beta]{2}} \, , 0 \right)$. We remark that $g^1 (v)=g (v)$, $\beta \to \infty$ implies $g^\beta ( v ) \to \left( 2,1,0 \right)$, and $\beta \downarrow 0$ implies $g^\beta ( v ) \to (3,0,0)$. 

We conclude that there is a close relationship between these generalised Gately solutions and the Core of this game in the sense that every generalised Gately solution is in the Core and that the relative interior of the Core is mapped through these generalised Gately solutions:
\[
\left\{ g^\beta ( v ) \mid \beta >0 \right\} = C^o(v) = \{ (t,3-t,0) \mid 2 < t <3 \} \varsubsetneq C(v) .
\]
This close relationship between these values and the Core refers directly to \citet{Gately1974}'s original motivation to identify his solutions as Core selectors. This is explored further below.

\section{Cooperative games and Gately values}

We first discuss the foundational concepts of cooperative games and solution concepts. Let $N = \{ 1, \ldots , n \}$ be an arbitrary finite set of players and let $2^N = \{ S \mid S \subseteq N \}$ be the corresponding set of all (player) coalitions in $N$. For ease of notation we usually refer to the singleton $\{ i \}$ simply as $i$. Furthermore, we use the simplified notation $S-i = S \setminus \{ i \}$ for any $S \in 2^N$ and $i \in S$. 

A \emph{cooperative game} on $N$ is a function $v \colon 2^N \to \mathbb R$ such that $v ( \varnothing ) =0$. A game assigns to every coalition a value or ``worth'' that this coalition can generate through the cooperation of its members. We refer to $v(S)$ as the \emph{worth} of coalition $S \in 2^N$ in the game $v$. The class of all cooperative games in the player set $N$ is denoted by
\[
\mathbb V^N = \{ v \mid v \colon 2^N \to \mathbb R \mbox{ such that } v ( \varnothing ) =0 \} .
\]
For every player $i \in N$ let $v_i = v ( \{ i \} )$ be her individually feasible worth in the game $v$. We refer to the game $v$ as being \emph{zero-normalised} if $v_i =0$ for all $i \in N$. The collection of all zero-normalised games is denoted by $\mathbb V^N_0 \subset \mathbb V^N$. 

The set $\mathbb V^N$ is a $(2^n-1)$-dimensional Euclidean vector space. For the analysis of games it useful to use the \emph{unanimity basis} of this Euclidean vector space. Here, for every coalition $\varnothing \neq S \subseteq N$ the $S$-\emph{unanimity game} $u_S \in \mathbb V^N$ is defined by
\begin{equation}
	u_S (T) = \left\{
	\begin{array}{ll}
		1 & \mbox{if } S \subseteq T \\
		0 & \mbox{otherwise}
	\end{array}
	\right.
\end{equation}
Every game $v \in \mathbb V^N$ can now be written as $v = \sum_{S \neq \varnothing} \Delta_S (v) \, u_S$, where $\Delta_S (v) = \sum_{T \subseteq S} (-1)^{|S|-|T|} \, v(T)$ is the Harsanyi dividend \citep{Harsanyi1974} of coalition $S$ in game $v$.

\paragraph{Marginal contributions and related classes of games}

The \emph{marginal contribution} of an individual player $i \in N$ in the game $v \in \mathbb V^N$ is defined by her marginal or ``separable'' contribution to the grand coalition in this game, i.e., $M_i (v) = v(N) - v(N-i)$ where $N-i = N \setminus \{ i \}$. This marginal contribution can be considered as a ``utopia'' value \citep{Tijs1981,Tijs2008} for the following classes of cooperative games:
\begin{definition}
	A cooperative game $v \in \mathbb V^N$ is 
	\begin{itemize}
		\item \textbf{essential} if it holds that
	\begin{equation}
		\sum_{j \in N} v_j \leqslant v(N) \leqslant \sum_{j \in N} M_j (v)
	\end{equation}
	\item \textbf{semi-standard} if for every player $i \in N$ it holds that
	\begin{equation}
		v_i \leqslant M_i (v) \quad \mbox{or, equivalently,} \quad v_i + v(N-i) \leqslant v(N)
	\end{equation}
	\item \textbf{semi-regular} if $v$ is essential as well as semi-standard.
	\item \textbf{standard} if $v$ is semi-standard and, additionally, for at least one player $j \in N$ it holds that $v_j < M_j (v)$, or, equivalently, $v_j + v(N-j) < v(N)$.
	\item \textbf{regular} if $v$ is essential as well as standard. The collection of regular cooperative games is denoted by $\mathbb V^N_{\star} \subset \mathbb V^N$.\footnote{We emphasise that every regular game $v \in \mathbb V^N$ satisfies a partitional form of superadditivity in the sense that $v(N-i) + v_i \leqslant v(N)$ for every $i \in N$, which is aligned with the notion of a game being \emph{weak constant-sum} as defined in \citet[Definition 5]{Gately2019}. Furthermore, \citet[Theorem 1(a)]{Gately2019} is also founded on the class of regular cooperative games.}
	\end{itemize}
\end{definition}
The class of regular cooperative games $\mathbb V^N_\star$ is the main domain of analysis for various forms of Gately solutions and their generalisations. In particular, we denote the collection of regular zero-normalised games by $\widehat{\mathbb V}^N = \mathbb V^N_{\star} \cap \mathbb V^N_0$.

An \emph{allocation} in the game $v \in \mathbb V^N$ is any point $x \in \mathbb R^N$ such that $x(N)=v(N)$, where we denote by $x(S) = \sum_{j \in S} x_j$ the allocated payoff to the coalition $S \in 2^N$.  We denote the class of all allocations for the game $v \in \mathbb V^N$ by $\mathbb A (v) = \{ x \in \mathbb R^N \mid x(N) = v(N) \} \neq \varnothing$. We emphasise that allocations can assign positive as well as negative payoffs to individual players in a game.

An \emph{imputation} in the game $v \in \mathbb V^N$ is an allocation $x \in \mathbb A(v)$ that is individually rational in the sense that $x_i \geqslant v_i$ for every player $i \in N$. The corresponding imputation set of $v \in \mathbb V^N$ is now given by $\mathbb I (v) = \{ x \in \mathbb A(v) \mid x_i \geqslant v_i$ for all $i \in N \}$. We remark that for any essential game $v$ with $v(N) > \sum_{i \in N} v_i$ the imputation set $\mathbb I(v)$ is a polytope with a non-empty interior. In particular, this holds for the class of regular games $\mathbb V^N_\star$.

We recall that for any cooperative game $v \in \mathbb V^N$, the \emph{Core} is defined as a set of imputations $C(v) \subset \mathbb I (v)$ such that $x \in C(v)$ if and only for all coalitions $S \in 2^N \colon x(S) \geqslant v(S)$. Hence,
\begin{equation}
	C(v) = \{ x \in \mathbb I (v) \mid x(S) \geqslant v(S) \mbox{ for all } S \in 2^N \, \}.
\end{equation}
Let $\mathbb V \subseteq \mathbb V^N$ be some collection of TU-games on player set $N$. A \emph{value} on $\mathbb V$ is a map $\phi \colon \mathbb V \to \mathbb R^N$ such that $\phi (v) \in \mathbb A (v)$ for every $v \in \mathbb V$. We emphasise that a value satisfies the efficiency property that $\sum_{i \in N} \phi_i (v) = v(N)$ for every $v \in \mathbb V$. We remark that a value $\phi$ is \emph{individually rational} (IR) if $\phi (v) \in \mathbb I (v)$ for all $v \in \mathbb V$.

\subsection{Gately points and Gately values}

\citet{Gately1974} seminally introduced a specific methodology to identify outcomes of a bargaining process that is different from the well-known notions of other bargaining solutions such as the Bargaining Set, the Kernel \citep{DavisMascher1965} and the Nucleolus \citep{Schmeidler1969}. Gately's approach is based on the notion of the ``propensity to disrupt''.
\begin{definition} \emph{\citep{Gately1974,Gately1976}} \\
	Let $v \in \mathbb V^N$ be a cooperative game on $N$. The \textbf{propensity to disrupt} of the coalition $S \in 2^N$ at allocation $x \in \mathbb A(v)$ is defined by
	\begin{equation}
		d(S,x) = \frac{x(N \setminus S) - v(N \setminus S)}{x(S) -v(S)}
	\end{equation}
	The \textbf{propensity to disrupt of player} $i \in N$ at allocation $x \in \mathbb A (v)$ is given by
	\begin{equation}
		d_i (x) = d ( \{ i \} ,x) = \frac{x(N-i) - v(N-i)}{x_i -v_i} = \frac{M_i (v) - x_i}{x_i - v_i} = \frac{M_i (v) - v_i}{x_i-v_i} -1
	\end{equation}
	A \textbf{Gately point} of the game $v \in \mathbb V^N$ is defined as an imputation $g \in \mathbb I(v)$ that minimises the individual propensities to disrupt, i.e., for all players $i \in N \colon$
	\begin{equation}
		d_i (g) \leqslant \min_{x \in \mathbb I(v)} \, \max_{j \in N} d_j(x)
	\end{equation}
\end{definition}
Gately points of cooperative games have most recently been explored by \citet{Gately2019}. They showed the following properties.\footnote{A proof of the properties collected here can be found in \citet{Gately2019}.}
\begin{lemma} \label{prop:GatelyOriginal} \emph{\citep{Gately2019}}
	\begin{abet}
		\item Every standard cooperative game $v \in \mathbb V^N$ admits a unique Gately point $g(v) \in \mathbb I(v)$ given by
		\begin{equation} \label{eq:GatelyDef}
			g_i (v) = v_i + \frac{M_i (v) - v_i}{\sum_{j \in N} \left( \, M_j (v) - v_j \, \right)} \left( \, v(N) - \sum_{j \in N} v_j \, \right)
		\end{equation}
		for every $i \in N$.
		\item For every standard zero-normalised game $v \in {\mathbb V}^N$ the unique Gately point is given by
		\begin{equation}
			g (v) =\frac{v(N)}{\sum_{j \in N} M_j (v)} \, M(v) \in \mathbb I(v) .
		\end{equation}
	\end{abet}
\end{lemma}

\noindent
Lemma \ref{prop:GatelyOriginal}(a) allows us to introduce the \textbf{Gately value} as the map $g \colon \mathbb V^N_\star \to \mathbb R^N$ on the class of regular cooperative games defined by equation (\ref{eq:GatelyDef}).

We emphasise that the Gately value is only non-trivially defined on the class of regular cooperative games $\mathbb V^N_\star$, while Gately points are in principle defined for arbitrary cooperative games with the property that $M_i(v) \neq v_i$ for some $i \in N$. As pointed out by \citet{Gately2019}, there might be games that admit no Gately points and other games that might admit multiple Gately points.

\paragraph{Generalised Gately values}

We generalise the notion of Gately points as seminally introduced in \citet{Gately1974}. The next definition introduces a generalised notion of the Gately value on the class of standard cooperative games. 
\begin{definition}
	Let $v \in \mathbb V^N$ be some standard cooperative game on $N$. For any parameter value $\alpha > 0$ we define the \textbf{$\alpha$-Gately value} as the imputation $g^\alpha (v) \in \mathbb I(v)$ with
	\begin{equation}
		g^\alpha_i (v) = v_i + \frac{\left( M_i (v) - v_i \right)^\alpha}{\sum_{j \in N} \left( M_j (v) - v_j \right)^\alpha} \, \left( \, v(N) - \sum_{j \in N} v_j \, \right)  \qquad \mbox{for every $i \in N$.}
	\end{equation}
	We refer to $\mathcal G = \{ g^\alpha \colon \mathbb V^N_\star \to \mathbb R^N \mid \alpha > 0 \}$ as the family of generalised Gately values on the domain of regular cooperative games $\mathbb V^N_\star$. For any $v \in \mathbb V^N_\star$ the related set $\mathcal G (v) = \{ g^\alpha (v) \mid \alpha >0 \} \subseteq \mathbb I(v)$ defines the \textbf{Gately set} for that particular game.
\end{definition}
From this definition we can identify some special cases:
\begin{itemize}
	\item We note that $g^1 = g \in \mathcal G$ is the original Gately value on the class of regular games $\mathbb V^N_\star$.
	\item Although $g^\alpha$ is not defined for $\alpha =0$, note that 
	\[
	\lim_{\alpha \downarrow 0} g^\alpha_i (v) = v_i + \tfrac{1}{|N_0 (v)|} \left( \, v(N) - \sum_{j \in N} v_j \, \right)
	\]
	for all $i \in N_0 (v)$ and $\lim_{\alpha \downarrow 0} g^\alpha_i (v) = v_i$ for $i \in N \setminus N_0 (v)$, where $N_0 (v) = \{ i \in N \mid M_i (v) > v_i \} \neq \varnothing$. This compares to the CIS value \citep{DriessenFunaki1991}. \\ Furthermore, if the game $v \in \widehat{\mathbb V}^N$ is additionally zero-normalised, $\lim_{\alpha \downarrow 0} g^\alpha_i (v)$ corresponds to the equal division value given by $E(v) = \tfrac{v(N)}{n}$ if $M_i (v) \neq 0$ for all $i \in N$.
	\item Finally, $\lim_{\alpha \to \infty} g^\alpha_i (v) = v_i + \tfrac{1}{|N_1 (v)|} \left( \, v(N) - \sum_{j \in N} v_j \, \right)$ for all $i \in N_1(v)$ and $\lim_{\alpha \to \infty} g^\alpha_i (v) = v_i$ for $i \in N \setminus N_1(v)$, where $N_1 (v) = \{ i \in N \mid M_i (v) - v_i = \max_{j \in N} (M_j (v) - v_j) \, \} \neq \varnothing$. Again, this can be interpreted as a variation of the CIS value.
\end{itemize}
The next definition introduces a generalised formulation of Gately's seminal notion of the propensity to disrupt. We show that $\alpha$-Gately values are closely related to optimisation problems based on this generalised notion.
\begin{definition}
	Let $v \in \mathbb V^N$ be some cooperative game on player set $N$. For every parameter $\beta >0$ the corresponding \textbf{generalised $\beta$-propensity to disrupt of player} $i \in N$ at imputation $x \in \mathbb I (v)$ is defined by
\begin{equation}
	\rho^\beta_i (x) = \frac{M_i (v) - v_i}{\left( x_i - v_i \right)^\beta}
\end{equation}
\end{definition}
We note here that for $\beta =1$, this generalised propensity to disrupt corresponds exactly to the original propensity to disrupt for an individual player as introduced by \citet{Gately1974}, in the sense that $\rho^1_i (x) = \frac{M_i (v) -v_i}{x_i-v_i} = d_i (x)+1$.

The following theorem shows the relationship between the balancing of such generalised propensities to disrupt and corresponding $\alpha$-Gately values. In particular, it is shown that the $\alpha$-Gately value can be interpreted as a bargaining value, like the original Gately value and the Nucleolus. Furthermore, for certain values of $\alpha$, the minimisation of the generated total generalised propensity to disrupt at an allocation results in the corresponding $\alpha$-Gately value.
\begin{theorem} \label{thm:MainExist}
	Let $v \in \mathbb V^N_{\star}$ be a regular cooperative game on $N$.
	\begin{abet}
		\item Let $\alpha >0$ and define $\beta = \tfrac{1}{\alpha}$. Then the $\alpha$-Gately value $g^\alpha (v) \in \mathbb I(v)$ is the unique $\beta$-Gately point in the sense that $g^\alpha (v)$ is the unique imputation that satisfies the property that
		\begin{equation} \label{eq:GatelyBalance}
			\rho^\beta_i \left( g^\alpha (v) \, \right) \leqslant \min_{x \in \mathbb I(v)} \, \max_{j \in N} \, \rho^\beta_j (x)
		\end{equation}
		for every player $i \in N$.
		\item Let $0< \alpha <1$ and define $\beta = \tfrac{1- \alpha}{\alpha} >0$. Then the $\alpha$-Gately value $g^\alpha (v) \in \mathbb I(v)$ is the unique solution to the minimisation of the total aggregated generalised $\beta$-propensity to disrupt of the game $v \colon$
			\begin{equation} \label{eq:GatelyMin}
				g^\alpha (v) = {\arg \min}_{x \in \mathbb I(v)} \, \sum_{j \in N} \rho^\beta_j (x)
			\end{equation}
	\end{abet}
\end{theorem}
\begin{proof}
To show assertion (a), we note that for every imputation $x \in \mathbb I(v)$ and every player $i \in N \colon \rho^\beta_j (x) = \frac{M_i (v) - v_i}{(x_i - v_i )^\beta} \geqslant 0$ from the hypothesis that $M_i(v) \geqslant v_i$. Furthermore, since $M_j (v) > v_j$ for at least one $j \in N$, we conclude that $r = \max_{j \in N} \rho^\beta_j (x) >0$. \\
Hence, we conclude that the minimax problem $\min_{x \in \mathbb I(v)} \, \max_{j \in N} \, \rho^\beta_j (x)$ can be solved by identifying $r >0$ and an imputation $x^* \in \mathbb I(v)$ such that $\rho^\beta_i (x^* ) = r$ for all $i \in N_0 = \{ i \in N \mid M_i (v) > v_i \} \neq \varnothing$. \\
First, note that for all $j \in N \setminus N_0 = \{ j \in N \mid M_j(v) = v_j \}$ we can set $x_j = v_j$. Next, for $i \in N_0$ we can now solve for $r >0$ as well as $x_i$. Rewriting $\rho^\beta_i (x) =r$, we derive that $x_i = \left( \frac{M_i(v)-v_i}{r} \right)^{\tfrac{1}{\beta}} + v_i$. \\
Note that, since $M_i(v) \geqslant v_i$ for all $i \in N$, it follows that $x \in \mathbb I(v)$. Hence,
\[
\sum_{i \in N} x_i = \frac{\sum_{i \in N_0} (M_i(v)-v_i)^{\tfrac{1}{\beta}}}{r^{\tfrac{1}{\beta}}} + \sum_{i \in N} v_i \equiv v(N)
\]
Since $\sum_{i \in N_0} (M_i(v)-v_i)^{\tfrac{1}{\beta}} = \sum_{i \in N} (M_i(v)-v_i)^{\tfrac{1}{\beta}}$, we conclude that
\[
r = \frac{\left[ \sum_{i \in N} (M_i (v) - v_i)^{\tfrac{1}{\beta}} \, \right]^\beta}{\left[ \, v(N) - \sum_{j \in N} v_j \, \right]^\beta} >0.
\]
From this we conclude that the identified solution exists and is unique under the regularity conditions on the game $v$. \\
Substituting the formulated solution of $r$ back into the formulation for the solution, we deduce that
\[
x_i = v_i + \frac{(M_i (v) - v_i)^{\tfrac{1}{\beta}}}{\sum_{j \in N} (M_j (v) - v_j)^{\tfrac{1}{\beta}}} \, \left( v(N) - \sum_{j \in N} v_j \right) \geqslant v_i .
\]
Recalling that $\beta =\tfrac{1}{\alpha}$, we indeed conclude that $x_i = g^\alpha_i (v)$, leading us to conclude that assertion (a) has been shown.

\medskip\noindent
To show assertion (b), consider the minimisation problem $\min_{ x \in \mathbb I(v)} R^\beta (x)$ as formulated, where $R^\beta (x) = \sum_{j \in N} \rho^\beta_j (x)$. Deriving the Lagrangian $L(x_1,...,x_n,\lambda)=  \sum_{i \in N} \left[ \,  \frac{M_i-v_i}{(x_i-v_i)^{\beta}} \, \right]+ \lambda( \sum_{i \in N} x_i - v(N))$, and deriving the necessary first-order conditions, we conclude that
\[
 \frac{M_1-v_1}{(x_1-v_1)^{\beta+1}} = \frac{M_2-v_2}{(x_2-v_2)^{\beta+1}} =  \cdots =   \frac{M_n-v_n}{(x_n-v_n)^{\beta+1}}.
 \]
 Thus, we arrive at $n-1$ equations given by
 \[
 x_k-v_k= \frac{(M_2-v_2)^{\frac{1}{\beta+1}} }{  (M_1-v_1)^{\frac{1}{\beta+1}}} (x_1-v_1) \qquad \mbox{for } k=2, \ldots ,n.
 \]
 This we can rewrite as
 \[
 v(N)- \sum_{j=3}^n x_j-x_1-v_2= \frac{(M_2-v_2)^{\frac{1}{\beta+1}} }{  (M_1-v_1)^{\frac{1}{\beta+1}}} (x_1-v_1)
 \]
 together with
 \[
 x_k=v_k+ \frac{(M_k-v_k)^{\frac{1}{\beta+1}} }{  (M_1-v_1)^{\frac{1}{\beta+1}}} (x_1-v_1) \qquad \mbox{for } k=3, \ldots ,n.
 \]
 Summing up the LHSs and the RHSs, we have the following equalities:
 \[
 v(N)-  x_1-v_2= v_3+...+v_n +  \sum_{j=2}^n   \frac{(M_j-v_j)^{\frac{1}{\beta+1}} }{  (M_1-v_1)^{\frac{1}{\beta+1}}} (x_1-v_1)
 \]
 This leads to the conclusion that
 \begin{align*}
 	 v(N)-   \sum_{j=2}^n  v_j  &  = x_1+   \frac{ (x_1-v_1)}{  (M_1-v_1)^{\frac{1}{\beta+1}}}   \sum_{j=2}^n  (M_j-v_j)^{\frac{1}{\beta+1}} \\
 	 &  =  x_1+   \frac{ (x_1-v_1)}{  (M_1-v_1)^{\frac{1}{\beta+1}}}   \sum_{j=1}^n  (M_j-v_j)^{\frac{1}{\beta+1}} -  (x_1-v_1)
 \end{align*}
 Hence, we conclude that
 \[
 \frac{ (M_1-v_1)^{\frac{1}{\beta+1}} }{ \sum_{j=1}^n  (M_j-v_j)^{\frac{1}{\beta+1}} }  [v(N)-   \sum_{j=1}^n  v_j ]  =    x_1-v_1 .
 \]
 Remarking that $\alpha = \tfrac{1}{\beta +1}$ leads us immediately to the insight that the first player's allocation is actually her $\alpha$-Gately value value. The resulting allocations for the other players $j= 2, \ldots ,n$ are derived in a similar fashion.
\end{proof}

\medskip\noindent
We remark that Theorem \ref{thm:MainExist} applies to regular cost games or problems as well. Indeed, for a \emph{cost game} $ v \in \mathbb V^N$ satisfying $v(N) \leqslant \sum_{j \in N} v_J$, $M_i (v) \leqslant v_i \leqslant 0$ for all $i \in N$ and $M_j (v)< v_j \leqslant 0$ for some $j \in N$, both assertions of Theorem \ref{thm:MainExist} hold. We do not consider these games here, but refer to, e.g., \citet{Moulin2004} for a discussion of these cost games.

To illustrate the importance of regularity of those cooperative games for which Gately values are well-defined as imposed in Theorem \ref{thm:MainExist}(a), we consider the next example of a three-player game that exhibits non-regularities.
\begin{example}
	Consider a 3-player game $v$ on $N = \{ 1,2,3 \}$ defined by $v_1=2$, $v_2=1$, $v_3=0$, $v(12) = v(13) = v(23) =4$ and $v(N)=5$. \\ 
	We remark that the marginal contributions can now be computed as $M_1 = M_2 = M_3=1$, leading to the conclusion that $M_1 - v_1 =-1 <0$, $M_2 = v_2$, and $M_3 - v_3 =1 >0$. Hence, this game is clearly neither essential nor semi-standard.\footnote{We also remark that this game has actually an empty Core.} \\
	We show that this game admits a continuum of Gately points, thereby illustrating that this game does not have a well-defined, unique Gately value. Note that for any proposed solution $x \in \mathbb A (v) \colon$
	\[
	\rho_1 (x) = \frac{-1}{x_1-2} \qquad \rho_2 (x) =0 \qquad \rho_3 (x) = \frac{1}{x_3}
	\]
	All Gately points are now characterised by two equations, namely $x(N) = v(N)$ and $\rho_1 (x) = \rho_3 (x) \geqslant 0$. This leads to the conclusion that the set of Gately points is a continuum given by $\{ (t,3,2-t) \mid 0 \leqslant t \leqslant 2 \} \subset \mathbb A(v)$. Note that this set of Gately points includes allocations that are \emph{not} imputations.
\end{example}
	
\noindent
With regard to Theorem \ref{thm:MainExist}(b) we remark that for $\beta =1$ the minimisation of the aggregated total propensity to disrupt $R^1 (x) = \sum_{i \in N} d_i(x)$ results in the $\tfrac{1}{2}$-Gately value as the unique solution. Furthermore, if $\beta =0$, the generalised propensity to disrupt for any player $i \in N$ is no longer a function of the allocation $x_i$, implying that the total aggregated $0$-propensity to disrupt is a constant function. This implies that the minimisation problem (\ref{eq:GatelyMin}) has a continuum of solutions, including all Gately points. 

\subsection{Dual Gately values}

Let $v \in \mathbb V^N$ be a cooperative game. Then the \emph{dual game} of $v$, denoted by $v^* \colon 2^N \to \mathbb R$, is defined by
\begin{equation}
	v^* (S) = v(N) - v(N \setminus S) \qquad \mbox{for every coalition } S \in 2^N
\end{equation}
The dual of a game assigns to every coalition $S \subseteq N$ the worth that is lost by the grand coalition $N$ if coalition $S$ leaves the game. Note in particular that $v^* ( \varnothing ) =0$, $v^* (N) = v(N)$ and $v^*_i = v^* ( \{ i \} ) = v(N) - v(N-i) = M_i (v)$ for all $i \in N$. Finally, $M_i (v^*) = v_i$ for every $i \in N$.

We investigate the ``dual'' of a given value, which assign to games the value of its dual game. As an illustrative example, we note that \citet{DriessenFunaki1991} considered the dual of the CIS-value, defined as the CIS-value of the dual game. They refer to this notion as the ``Egalitarian Non-Separable Contribution'' value, or ENSC-value. 

We can apply a similar procedure to the Gately value. We note first that the dual of a Gately value only can properly formulated for parameter values that are natural numbers, i.e., $\alpha \in \mathbb N$. This is subject to the next definition.

\begin{definition}
	Let $\alpha \in \mathbb N$. The \textbf{dual $\alpha$-Gately value} is a map $\overline{g^\alpha} \colon \mathbb V^N_\star \to \mathbb R^N$ that assigns to every regular cooperative game $v \in \mathbb V^N_\star$ the $\alpha$-Gately value of its dual game $v^* \in \mathbb V^N$, i.e., $\overline{g^\alpha} (v) = g^\alpha (v^*) \in \mathbb A(v)$.
\end{definition}

\noindent
The next proposition considers some properties of dual $\alpha$-Gately values. 

\begin{proposition} \label{prop:DualGately}
	Consider a regular cooperative game $v \in \mathbb V^N_{\star}$ and let $\alpha \in \mathbb N$ be a natural number. Then the following properties hold:
	\begin{abet}
		\item For every $\alpha \in \mathbb N$ the dual $\alpha$-Gately value of $v$ is well-defined and given by
		\begin{equation}
			\overline{g^\alpha_i} (v) = M_i (v) - \frac{(M_i(v) - v_i)^\alpha}{\sum_{j \in N} (M_j(v) - v_j)^\alpha} \left( \, \sum_{j \in N} M_j (v) - v(N) \, \right)
		\end{equation}
		for every player $i \in N$.
		\item The dual $\alpha$-Gately value of $v$ is identical to the $\alpha$-Gately value of $v$, i.e., $\overline{g^\alpha} (v) = g^\alpha (v)$, if and only if $\alpha =1$ and/or $M_i (v) - v_i = M_j (v) -v_j \geqslant 0$ for all $i,j \in N$.
	\end{abet}
\end{proposition}

\begin{proof}
	To show assertion (a), let $\alpha \in \mathbb N$. We compute that for every player $i \in N \colon$
	\begin{align*}
		\overline{g^\alpha_i} (v) & = g^\alpha_i (v^*) = v^*_i + \frac{(M_i (v^*) - v^*_i)^\alpha}{\sum_{j \in N} (M_j (v^*) - v^*_j)^\alpha} \, \left( \, v^*(N) - \sum_{j \in N} v^*_j \, \right) \\[1ex]
		& = M_i (v) + \frac{(v_i - M_i (v) \, )^\alpha}{\sum_{j \in N} (v_j - M_j (v) \, )^\alpha} \, \left( \, v(N) - \sum_{j \in N} M_j (v) \, \right) \\[1ex]
		& = M_i (v) - \frac{(-1)^\alpha \, (M_i (v) - v_i )^\alpha}{(-1)^\alpha \, \sum_{j \in N} (M_j (v) - v_j )^\alpha} \, \left( \, \sum_{j \in N} M_j (v) - v(N) \, \right) \\[1ex]
		& = M_i (v) - \frac{(M_i (v) - v_i )^\alpha}{\sum_{j \in N} (M_j (v) - v_j )^\alpha} \, \left( \, \sum_{j \in N} M_j (v) - v(N) \, \right)
	\end{align*}
	with the remark that $(-1)^\alpha$ attains only the values $1$ and $-1$ due to $\alpha \in \mathbb N$. \\
	Furthermore, we note that $\sum_{i \in N} \overline{g^\alpha_i} (v) = v(N)$, thereby showing that the dual $\alpha$-Gately value is indeed well-defined. 
	
	\medskip\noindent
	To show assertion (b) let $i \in N$ and $\alpha \in \mathbb N$.  We now note that from assertion (a) $g^\alpha_i (v) = \overline{g^\alpha_i} (v)$ if and only if
	\[
	v_i + \frac{(M_i(v) - v_i)^\alpha}{\sum_{j \in N} (M_j(v) - v_j)^\alpha} \left( \, v(N) - \sum_{j \in N} v_j \, \right) = M_i (v) - \frac{(M_i(v) - v_i)^\alpha}{\sum_{j \in N} (M_j(v) - v_j)^\alpha} \left( \, \sum_{j \in N} M_j (v) - v(N) \, \right)
	\]
	or
	\[
	\frac{\sum_{i \in N} \, (M_i(v) - v_i)^\alpha}{\sum_{j \in N} \, (M_j(v) - v_j)} = (M_i(v) - v_i)^{\alpha -1}
	\]
	This is valid for all $i \in N$ if and only if $\alpha =1$ and/or $M_i (v) - v_i = M_j (v) - v_j \geqslant 0$ for all $i,j \in N$.
\end{proof}

\bigskip\noindent
Proposition \ref{prop:DualGately} (b) implies immediately that the dual Gately value is the same as the Gately value on the class of regular games. This is stated in the next corollary.
\begin{corollary}
	For every regular cooperative game $v \in \mathbb V^N_{\star}$, the dual Gately value of $v$ is identical to the Gately value of $v$, i.e., $\overline{g}_i (v) = g_i (v)$ for all $i \in N$.
\end{corollary}

\subsection{Constructive characterisations of the Gately value}

It is easy to see that the Gately value on the class of regular games $\mathbb V^N_\star$ is a compromise value of the individual worth vector and the net marginal contribution vector. Indeed, for any regular game $v \in \mathbb V^N_\star$, the individual worth vector is given by $\nu_v = ( v_1, \ldots , v_n )$ and the net marginal contribution vector by $\bar n = M(v) - \nu_v = ( M_1 (v) - v_1, \ldots , M_n (v)-v_n \, )$. Now the Gately value $g(v)$ for game $v$ can be written as
\[
g(v) = \nu_v + \gamma_v \cdot \left( M(v) - \nu_v \right) = (1- \gamma_v \, ) \nu_v + \gamma_v \, M(v)
\]
where
\[
\gamma_v = \frac{v(N) - \sum_{i \in N} v_i}{\sum_{i \in N} ( M_i (v) - v_i )} .
\]
This re-interpretation of the Gately value as a compromise value holds on the class of regular games $\mathbb V^N_\star$ and allows us to characterise the Gately value as such a compromise value.

We remark that the compromise values form a distinct subclass of solution concepts that are based on a similar methodology of determining the exact allocation of the worth of the grand coalition $N$. This refers to the fundamental property that the value is a convex combination of a well-defined lower and upper bound such that the value satisfies efficiency. We refer to \citet{TijsOtten1993} and \citet{GillesBrink2023} for detailed discussion and characterisations of these solution concepts.

\paragraph{A constructive axiomatisation of the Gately value}

\citet{Tijs1987} devised a simple axiomatisation for the $\tau$-value \citep{Tijs1981} that is completely based on the property that the $\tau$-value is a compromise value. We can devise a similar axiomatisation of the Gately value by replicating Tijs's characterisation methodology to the Gately value to arrive at the following axiomatisation. 

In this characterisation, a variant of the compromise property and the restricted proportionality property, seminally introduced by \citet{Tijs1987} on the class of quasi-balanced games, can be constructed on the class of regular cooperative games $\mathbb V^N_\star$.
\begin{theorem} \label{thm:Axiomatisation}
	The Gately value $g$ is the unique map $f \colon \mathbb V^N_\star \to \mathbb R^N$ on the class of regular games $\mathbb V^N_\star$ that satisfies the following three properties:
	\begin{numm}
		\item \textbf{Efficiency:} $\sum_{i \in N} f_i(v) = v(N)$ for every $v \in \mathbb V^N_\star$;
		\item \textbf{$\nu_v$-Compromise property:} For every regular game $v \in \mathbb V^N_\star \colon f(v) = \nu_v + f(v- \nu_v )$, where $v - \nu_v \in \widehat{\mathbb V}^N$ is the zero-normalisation of $v$ defined by $(v- \nu_v)(S) = v(S) - \sum_{i \in S} v_i$ for every coalition $S \in 2^N$, and;
		\item \textbf{Restricted proportionality property:} For every zero-normalised regular cooperative game $v \in \widehat{\mathbb V}^N \colon f(v) = \gamma_v \, M(v)$ for some $\gamma_v \in \mathbb R$.
	\end{numm}
\end{theorem}

\begin{proof}
	We first show that the Gately value $g \colon \mathbb V^N_\star \to \mathbb R^N$ satisfies the three stated properties. For that purpose let $v \in \mathbb V^N_\star$.
	\begin{numm}
		\item Obviously the Gately value $g(v)$ is efficient for $v$.
		\item Let $w=v- \nu_v \in \mathbb V^N_0$ be the zero-normalisation of $v$. Then for every $i \in N$ we deduce that $w_i = v_i - v_i =0$ and $M_i (w) = w(N) - w(N-i) = v(N) - v(N-i) - v_i = M_i (v) - v_i$. Hence, $M_i (w) \geqslant 0 = w_i$ and for those players $j \in N$ with $M_j (v) > v_j$ we deduce that $M_j (w) >0 = w_j$. \\
		Furthermore, $\sum_{i \in N} v_i \leqslant v(N) \leqslant \sum_{i \in N} M_i (v)$ is equivalent to $0 \leqslant v(N) - \sum_{i \in N} v_i = w(N) \leqslant \sum_{i \in N} M_i (v) - \sum_{i \in N} v_i = \sum_{i \in N} M_i (w)$, implying that $w \in \mathbb V^N_\star$. Therefore, $w = v- \nu_v \in \widehat{\mathbb V}^N$. \\
		Now by definition for every $i \in N \colon$
		\[
		g_i (w) = \frac{M_i (w)}{\sum_{j \in N} M_j (w)} \cdot w(N) = \frac{M_i (v)-v_i}{\sum_{j \in N} (M_j (v)-v_j)} \cdot \left( v(N) - \sum_{j \in N} v_j \right) = g_i(v) -v_i .
		\]
		This shows that $g_i (v) = v_i + g_i (v- \nu_v )$.
		\item Assume that $v \in \widehat{\mathbb V}^N$. Then for any $i \in N \colon g_i (v) = \frac{M_i (v)}{\sum_{j \in N} M_j (v)} \cdot v(N)$ showing restricted proportionality with $\gamma_v = \frac{v(N)}{\sum_{j \in N} M_j (v)}$.
	\end{numm}
	Next, we show that if $f \colon \mathbb V^N_\star \to \mathbb R^N$ satisfies the three stated properties, it is equal to the Gately value. Take any regular game $v \in \mathbb V^N_\star$ and let $w = v- \nu_v \in \widehat{\mathbb V}^N$ be its zero-normalisation. \\
	Then from restricted proportionality we have that $f(w) = \gamma_w M(w) = \gamma_w \left( M(v) - \nu_v \right)$. Furthermore, from the compromise property we conclude that
	\[
	f(v) = \nu_v + f (v - \nu_v) = \nu_v + \gamma_v ( M(v) - \nu_v) .
	\]
	Using efficiency we then conclude that
	\[
	\sum_{i \in N} f_i (v) = \sum_{i \in N} v_i + \gamma_v \left( \sum_{i \in N} M_i (v) - \sum_{i \in N} v_i \, \right) = v(N)
	\]
	implying that
	\[
	\gamma_v = \frac{v(N) - \sum_{i \in N} v_i}{\sum_{i \in N} (M_i (v) - v_i )} .
	\]
	We immediately conclude from this that $f_i (v) = g_i(v)$, showing the assertion.
\end{proof}

\bigskip\noindent
The axiomatisation in Theorem \ref{thm:Axiomatisation} is constructive in the sense that it shows basic properties satisfied by the Gately value. These three properties are independent as the next discussion shows.

The \emph{efficiency} property is a well-established property that is used throughout the literature. It guarantees that the allocation rule selects from the set of imputations in the game rather than the broader set of allocations. We note that the allocation rule $f(v)=M(v)$ on $\mathbb V^N_\star$ clearly satisfies the compromise property as well as the restricted proportionality property, but which is not efficient. 

The $\nu_v$-\emph{compromise property} is a reduced form of additivity and as such decomposes the allocation rule in a translation of the allocation assigned to the zero-normalisation of the game. This property originated in \citet{Tijs1987} as the ``compromise property'' for the minimal right vector $m(v)$ rather than the vector of individual worths $\nu_v$. It is clear that the $\tau$-value satisfies efficiency and the restricted proportionality property. It does not satisfy the $\nu_v$-compromise property, but rather the compromise property based on the minimal rights vector $m(v)$.

The \emph{restricted proportionality property} imposes zero-normalised games are assigned an allocation that is proportional to the utopia vector $M(v)$. This property originates from \citet{Tijs1987} as well and it is satisfied by the $\tau$-value. On the other hand, the Shapley value is a solution concept that is efficient and satisfies the $\nu_v$-compromise property, but it does not satisfy restricted proportionality. 

\section{Gately points and the Core for 3-player games}

\citet{Gately1974} introduced his solution concept as a Core selector within the setting of three-player games only, even though Gately did not investigate the exact conditions under which this solution is indeed in the Core. \citet{Gately1976} point out that Gately's conception does not necessarily result in a Core selector for games with more than three players, devising a counterexample for 4 players. 

In this section we first discuss the relationship between the Gately value and the Core for games with three players only. This is an exceptional case, since the worths of all coalitions in a three-player game are featured in the computation of the Gately value, in contrast to games with more than three players, in which worths of medium-sized coalitions are not considered. This is further explored in the second part of this section, which considers the relationship between the Gately value and the Core of cooperative games with an arbitrary number of players. 

We are able to confirm that there is a strong relationship between Gately points and the Core in three-player games. We first illustrate that there exist essential games with empty Cores for which the unique Gately point is well-defined. 

\begin{example}
	Consider an essential three-player game with $N = \{ 1,2,3 \}$ and $v$ given by $v_1=5$, $v_2=v_3=0$, $v(12) = v(13) =1$, $v(23)=5$ and $v(N)=6$. \\ First note that $v$ is indeed essential, since $M_1 (v) =1$ and $M_2 (v) = M_3 (v) =5$. On the other hand, $v$ is not semi-standard, since $v_1 =5 > M_1 (v) =1$.  \\ Note that the Core of this game is empty, since for an allocation $x \in \mathbb A (v)$ with $x(N) = v(N)=6$ and $x_2+ x_3 \geqslant v(23) = 5$ it follows that $x_1 \leqslant 1$. This is contradiction to the Core requirement that $x_1 \geqslant v_1= 5$. \\ Regarding the existence of Gately points for this particular game, we note that the minimax optimisation problem can be re-stated here as the balance equation $\frac{M_1 - v_1}{x_1-v_1} = \frac{M_2 - v_2}{x_2-v_2} = \frac{M_3 - v_3}{x_3-v_3}$ resulting into $\frac{-4}{x_1-5} = \frac{5}{x_2} = \frac{5}{x_3}$, which leads to a unique Gately point $g_1 = 4 \tfrac{1}{3}$ and $g_2 = g_3 = \tfrac{5}{6}$. Note that this unique Gately point can also be computed by the Gately value formula stated in equation (\ref{eq:GatelyDef}). \\ In comparison, the Shapley value of this game is given by $\phi = \left( \, 2 \tfrac{1}{3} \, , \, 1 \tfrac{5}{6} \, , \, 1 \tfrac{5}{6} \, \right)$.
\end{example}

\noindent
The next theorem gathers some properties of three-player games regarding the relationship between the Core and the Gately points of these games. These properties generalise the insights presented through the previous two examples. 

\begin{theorem} \label{thm:3playerCore}
	Let $v \in \mathbb V^N$ be a three-player game on $N = \{ 1,2,3 \}$. Then the following properties hold:
	\begin{abet}
		\item If the game $v$ is semi-regular, then the Gately value is in its Core, $g(v) \in C(v) \neq \varnothing$.
		\item If $C(v) \neq \varnothing$, then the game $v$ is semi-regular and $g(v) \in C(v)$. 
	\end{abet}
\end{theorem}

\begin{proof}
	To show assertion (a), we first consider a three-player game $v \in \mathbb V^N$ that is semi-regular, but not regular. Hence, $M_i = v_i$ for $i = 1,2,3$, implying that $v_1+v_2+v_3 = M_1(v) + M_2 (v) + M_3 (v) = v(N)$. Simple computations show that there is a unique Core imputation given by $C(v) = \left\{ \, \left( v_1,v_2,v_3 \right) \, \right\} = \left\{ \, \left( \, M_1(v) , M_2 (v) , M_3 (v) \, \right) \, \right\} \neq \varnothing$. Furthermore, it is easily established that the unique Gately point is well-defined and given by $g(v) = \left( v_1,v_2,v_3 \right) \in C(v)$. 
	
	\smallskip\noindent
	Next, we assume that $v$ is regular in the sense that $v_1+v_2+v_3 \leqslant v(N) \leqslant M_1 (v) + M_2(v) + M_3(v)$, $v_i \leqslant M_i(v)$ for all $i=1,2,3$ and, without loss of generality, $v_1 < M_1 (v)$. Hence, it holds that $v(12) + v(13) + v(23) \leqslant 2 v(N)$. Furthermore, it follows that $3v(N) - v(12)-v(13)-v(23)-v_1-v_2-v_3 = \sum_j \left( M_j(v) - v_j \, \right) >0$.
	\\
	Now define for every $i = 1,2,3$
	\[
	\eta_i = \frac{2v(N) - v(12)-v(13)-v(23)}{3v(N) - v(12)-v(13)-v(23)-v_1-v_2-v_3} \, \left(M_i (v) - v_i \, \right)
	\]
	Note that $\eta_i \geqslant 0$ for all $i=1,2,3$ and that, in particular, $\eta_1 >0$. \\
	We now note the following properties of these introduced quantities:
	\begin{itemize}
		\item First, regarding their sum,
		\begin{align*}
			\eta_1 +\eta_2 + \eta_3 & = \frac{2v(N) - v(12)-v(13)-v(23)}{3v(N) - v(12)-v(13)-v(23)-\sum_i v_i} \, \sum_i \left(M_i (v) - v_i \, \right) \\
			& = \sum_i M_i (v) - v(N) = 2v(N) - v(12)-v(13)-v(23) .
		\end{align*}
		\item Second, for every $i=1,2,3 \colon$
		\[
		\eta_i = \frac{\sum_j M_j (v) - v(N)}{\sum_j \left( M_j (v) - v_j \, \right)} \, \left(M_i (v) - v_i \, \right) \leqslant  M_i(v) -v_i .
		\]
		\item Finally, for every $i = 1,2,3$ we argue that $g_i (v) = M_i(v) - \eta_i$. Indeed,
		\begin{align*}
			g_i (v) & = \frac{M_i(v) - v_i}{\sum_j \left( M_j(v) - v_j \right)} \left( \, v(N) - \sum_j v_j \, \right) \\[1ex]
			& = \frac{\sum_j M_j (v) -v(N)}{\sum_j \left( M_j(v) - v_j \right)} \, v_i + \frac{v(N) -\sum_j v_j}{\sum_j \left( M_j(v) - v_j \right)} \, M_i (v) \\[1ex]
			& = M_i (v) + \frac{\sum_j M_j (v) - v(N)}{\sum_j \left( M_j(v) - v_j \right)} \, \left( v_i - M_i (v) \, \right) = M_i (v) - \eta_i .
		\end{align*}
	\end{itemize}
	Using the argument of \citet[4.12.1]{Vorobev1977}, we now claim that $g(v) \in C(v)$ using the construction above. We check the conditions for $g(v)$ being a Core selector:
	
	\smallskip\noindent
	First, for every $i = 1,2,3 \colon g_i(v) = M_i (v) - \eta_i \geqslant M_i (v) - \left( M_i (v) - v_i \, \right) = v_i$.
	\\
	Second, we can check for each 2-player coalition the Core conditions. For $\{1,2 \}$ it is easy to see that
	\begin{align*}
	g_1 (v) + g_2(v) & = M_1(v) + M_2(v) - \eta_1 - \eta_2 \\
	& = 2v(N) - v(23) - v(13)  - \eta_1 - \eta_2 \\
	& = v(12) + \eta_3 \geqslant v(12)
	\end{align*}
	Similar arguments show that $g_1 (v) + g_3(v) \geqslant v(13)$ and $g_2(v) + g_3 (v) \geqslant v(23)$. \\
	Together with $g_1 (v) + g_2(v) + g_3(v) = v(N)$, this completes the proof of assertion (a). 
	
	\medskip\noindent
	To show assertion (b), assume that for three-player game $v \in \mathbb V^N$ with $N= \{ 1,2,3 \}$ it holds that $C(v) \neq \varnothing$. Hence, there exists some $(x_1,x_2,x_3) \in \mathbb R^3$ with $x_1+x_2+x_3 = v(N)$, $x_i \geqslant v_i$ for $i=1,2,3$, and $x_1 + x_2 \geqslant v(12)$, $x_1 + x_3 \geqslant v(13)$, and $x_2 + x_3 \geqslant v(23)$.
	\\
	Adding the last three inequalities results in the conclusion that
	\[
	2v(N) = 2x_1+2x_2+2x_3 \geqslant v(12)+ v(13) + v(23),
	\]
	which in turn leads to the conclusion that
	\[
	M_1(v) + M_2(v) + M_3(v) = (v(N) - v(12)) + (v(N)-v(13) + (v(N)-v(23)) \geqslant v(N) .
	\]
	Furthermore, from $x_i \geqslant v_i$ for $i=1,2,3$ it follows that $v(N) = x_1+x_2+x_3 \geqslant v_1+v_2+v_3$. \\
	These two inequalities leads us to the conclusion that $v_1+v_2+v_3 \leqslant v(N) \leqslant M_1(v) + M_2 (v) + M_3 (v)$, implying that $v$ is indeed essential. \\
	Furthermore, $v(N) = x_1+(x_2+x_3) \geqslant v_1 + v(23)$ implying that $M_1 (v) = v(N) - v(23) \geqslant v_1$. This argument can be replicated for players 2 and 3, leading to the desired conclusion that $v$ is indeed semi-regular. The conclusion that, therefore, $g(v) \in C(v)$ follows from assertion (a).
\end{proof}

\bigskip\noindent
An immediate insight from Theorem \ref{thm:3playerCore} is that for every three-player game with a non-empty Core, the Gately value is a Core selector:
\begin{corollary}
	Let $v \in \mathbb V^N$ with $N = \{ 1,2,3 \}$ be a three-player cooperative game. Then $g(v) \in C(v)$ if and only $C(v) \neq \varnothing$. 
\end{corollary}
Similar arguments as the ones used in the proof of Theorem \ref{thm:3playerCore}(b) show that the Gately value of certain semi-regular three-player games is equal to the vector of marginal contributions and it is the unique Core imputation if the Core is non-empty.
\begin{corollary}  \label{corollary:3-playerCore}
	Let $v \in \mathbb V^N$ with $N = \{ 1,2,3 \}$ be a three-player cooperative game such that $v$ is an essential cooperative game such that $v(12)+v(13)+v(23)=2v(N)$. Then the unique Gately point coincides with the Nucleolus of the game, being equal to the vector of marginal contributions $g(v) = \mathcal N (v) = \left( \, M_1(v) , M_2(v) , M_3(v) \, \right)$. \\ Furthermore, if $C(v) \neq \varnothing$, it holds that $C(v) = \{ g(v) \} = \{ \mathcal N (v) \}$.
\end{corollary}
The conclusions about the equivalence of the Gately value and the Nucleolus of the three-player game stated in Corollary \ref{corollary:3-playerCore} follows from application of \citet[Theorem 1]{Parlar2010} in case that $C(v) = \varnothing$.

\paragraph{$\alpha$-Gately values and the Core of 3-player games}

The analysis of the relationship between $\alpha$-Gately values and the Core of a three-player game is more complex if we look beyond the standard Gately value ($\alpha =1)$. 

The next example shows that there exist three-player games in which $\alpha$-Gately values are in the Core for a certain closed interval of $\alpha$ values bounded away from zero.
\begin{example}
	Consider a zero-normalised three-player game $v$ with $N = \{ 1,2,3 \}$ and $v_i =0$ for $i = 1,2,3$, $v(12) = 12$, $v(13) = v(23)= 7$ and $v(N)=16$. Clearly, this game is regular. \\ We easily compute that the marginal contributions are given by $M_1 = M_2 =7$ and $M_3=4$. For any $\alpha > 0$ we compute the $\alpha$-Gately values as
	\[
		g^\alpha_1 (v) = g^\alpha_2 (v) = \frac{8 \cdot 7^\alpha}{7^\alpha + 2 \cdot 4^{\alpha -1}} \mbox{ and } g^\alpha_3 (v) = \frac{8 \cdot 4^\alpha}{7^\alpha + 2 \cdot 4^{\alpha -1}}
	\]
	We note that there are essentially two characteristic inequalities to determine whether the $\alpha$-Gately value in the Core of $v$:
	\[
		g^\alpha_1 (v) + g^\alpha_2 (v) \geqslant v(12)=12  \ \mbox{ and } \ g^\alpha_1 (v) + g^\alpha_3 (v) = g^\alpha_2 (v) + g^\alpha_3 (v) \geqslant v(13) = v(23) = 7
	\]
	The first inequality leads to the conclusion that $\alpha \geqslant \tfrac{\ln 3 - \ln 2}{\ln 7 - \ln 4} \approx 0.7245$ and the second inequality results in $\alpha \leqslant \tfrac{\ln 7 - \ln 2}{\ln 7 - \ln 4} \approx 2.2386$. Hence, the range of $\alpha$ values for which the $\alpha$-Gately value is in the Core of this game is given by $\alpha \in \left[ \, \tfrac{\ln 3 - \ln 2}{\ln 7 - \ln 4} , \tfrac{\ln 7 - \ln 2}{\ln 7 - \ln 4} \, \right]$. Note that $\alpha =1$ is indeed in this interval,  i.e., $g(v) \in C(v)$. \\ The Shapley value of this game is computed as $\phi = \left( 5 \tfrac{1}{2} \, , \, 5 \tfrac{1}{2} \, , \, 3 \right) = g^{\alpha^\bullet} (v) \in C(v)$ with $\alpha^{\bullet} = \tfrac{\ln 13 - \ln 6}{\ln 7 - \ln 4} \approx 1.38164$. \\ Finally, we remark that the Nucleolus of this game is given by $\mathcal N = (7,7,2) = g^{\alpha^\circ} (v) \in C(v)$ with $\alpha^{\circ} = \tfrac{\ln 7 - \ln 2}{\ln 7 - \ln 4}$, which is a corner point of the Core.
\end{example}
The next example considers a game with a large set of imputations and a minimal Core, consisting of a single imputation. For this example we show that the original Gately value is the only Core selector, while all $\alpha$-Gately values for $\alpha \neq 1$ are outside the Core.

\begin{figure}[h]
\begin{center}
\includegraphics[scale=0.25]{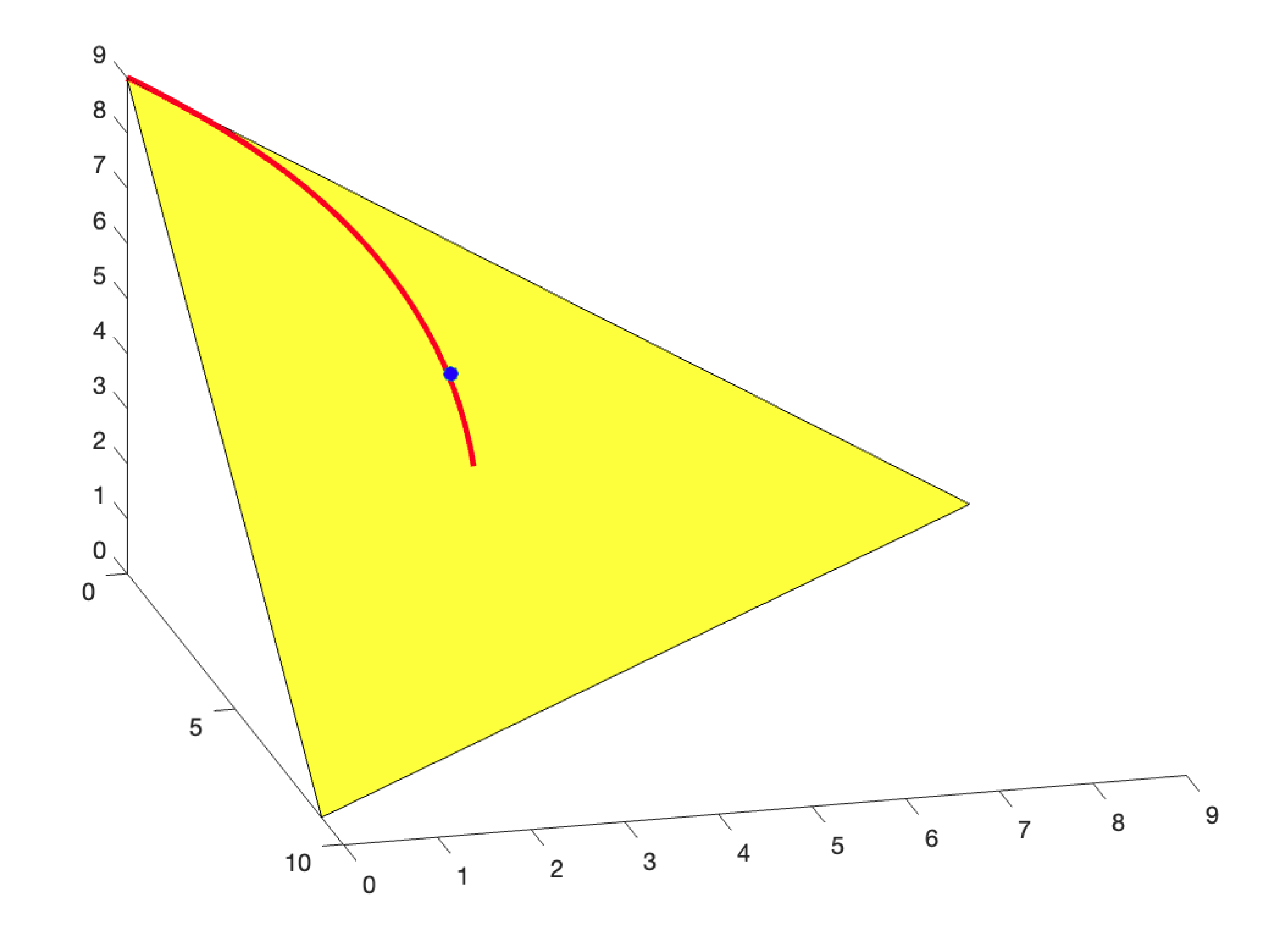}
\end{center}
\caption{The Core and the trajectory of $\alpha$-Gately values in Example \ref{ex:Example A}. }
\end{figure}

\begin{example} \label{ex:Example A}
	Consider a regular three-player game with $N = \{ 1,2,3 \}$ and $v$ given by $v_1= v_2=v_3=0$, $v(12) = 5$, $v(13) =6$, $v(23)=7$ and $v(N)=9$. Note that $\mathbb I(v) = \{ x \in \mathbb R^3_+ \mid \sum x_i =9 \}$ and that the Core is a singleton with $C(v) = \{ \,  (2,3,4) \, \} = \{ \, M(v) \, \}$. \\ We note that for this game $g(v) = M(v)$ selects the unique Core imputation. However, for all $\alpha >0$ with $\alpha \neq 1$ we have that
	\[
	g^\alpha (v) = \frac{9}{2^\alpha + 3^\alpha + 4^\alpha} \left( \, 2^\alpha , 3^\alpha , 4^\alpha \right) \neq (2,3,4) .
	\]
	Note that $g^\alpha (v) \to (3,3,3)$ as $\alpha \downarrow 0$ and $g^\alpha (v) \to (0,0,9)$ as $\alpha \to \infty$. This convergence is not monotone as one would possibly expect, since $g^\alpha_2 (v)$ attains a maximal value of $g^{\hat\alpha}_2 (v) \approx 3.0291$ at $\hat\alpha = \tfrac{1}{\ln 2} \, \ln \left[ \, \tfrac{\ln 3 - \ln 2}{\ln 4 - \ln 3} \, \right] \approx 0.4951$.
	\\
	The results of the analysis of this example are summarised in Figure 1. The yellow simplex represents the space of imputations $\mathbb I(v)$, while the Core is the unique imputation depicted as a blue point. The red curve denotes the set of $\alpha$-Gately values, $\{ g^\alpha (v) \mid \alpha >0 \}$. \\ Finally, we compute the Shapley value as $\phi = \left( 3 \tfrac{1}{2} \, , \, 3 \, , \,  2 \tfrac{1}{2} \right) \notin C(v)$, which cannot be expressed as a proper $\alpha$-Gately value. 
\end{example}

\section{Gately values of $n$-player games}

In this section we look at the relationship of Gately values and the Core as well as the Shapley value of arbitrary $n$-player games. We first consider the Gately value and its membership of the Core. Subsequently, we consider the relationship between the Shapley and Gately values. The next example illustrates the issues of the questions we investigate here.\footnote{We remark that \citet{Gately1976} already provided an example of a four-player game in which the Gately value is not in the non-empty Core of that game.}

\begin{example}
	Consider a regular zero-normalised four-player game $v$ with $N = \{ 1,2,3,4 \}$ and $v_i =0$ for all $i \in N$, $v(12) = 8$, $v(13) = v(14) = v(23) = v(24) = v(34) =1$, $v(123) = v(124) = 5$, $v(134) = v(234) =4$, and $v(N) =12$. We note that the Core of this game is non-empty, since $\left( 4 \tfrac{1}{2} \, , \, 4 \tfrac{1}{2} \, , \, 1 \tfrac{1}{2} \, , \, 1 \tfrac{1}{2} \right) \in C(v)$.
	\\
	From these worths, we derive that $M(v) = (4,4,3,3)$. From this it is easy to establish that the Gately value of this game is given by $g = \left( 3 \tfrac{3}{7} \, , \, 3 \tfrac{3}{7} \, , \, 2 \tfrac{4}{7} \, , \, 2 \tfrac{4}{7} \, \right)$. Clearly, $g \notin C(v)$ since $g_1+g_2 = 6 \tfrac{6}{7} < v(12) =8$. \\ Finally, we determine that the Shapley value of this game is given by $\phi =  \left( 3 \tfrac{3}{4} \, , \, 3 \tfrac{3}{4} \, , \, 2 \tfrac{1}{4} \, , \, 2 \tfrac{1}{4} \, \right) \notin C(v)$. Note that $g (v) \neq \phi (v)$.
\end{example}
In the following analysis we particularly focus on the conditions on $n$-player games under which $g (v) \in C(v)$ and/or $g(v) = \phi (v)$. We establish some full characterisations of these equivalences.

\subsection{Gately values and the Core}

The main condition for which a ``symmetric'' or ``anonymous'' cooperative game has a non-empty Core has been identified as the condition that for all coalitions $S \in 2^N \colon \tfrac{v(S)}{|S|} \leqslant \tfrac{v(N)}{n}$ \citep[page 149]{Shubik1982}. This condition has been referred to as ``domination by the grand coalition'' by \citet{Chatterjee1993} and as ``top convexity'' by \citet{JacksonNouweland2005}. We generalise this condition to identify when the $\alpha$-Gately value is in the Core of a regular, zero-normalised cooperative game.
\begin{definition}
	Let $v \in \mathbb V^N$ be a semi-standard cooperative game and let $\alpha > 0$. The cooperative game $v$ is said to be \textbf{$\alpha$-top dominant} if for every coalition $S \in 2^N$
	\begin{equation} \label{eq:TopDom}
		\left[ v(S) - \sum_{j \in S} v_j \right] \cdot \sum_{j \in N} \left( \, M_j (v) - v_j \, \right)^\alpha \leqslant \left[ v(N) - \sum_{j \in N} v_j \right] \cdot \sum_{j \in S} \left( \, M_j (v) - v_j \, \right)^\alpha .
	\end{equation}
	For $\alpha =1$ we refer to the started property as ``top dominance''.
\end{definition}
First we remark that $\sum_{j \in S} \left( \, M_j (v) - v_j \, \right)^\alpha \geqslant 0$ for every semi-standard cooperative game $v \in \mathbb V^N$ and every $\alpha > 0$.

Furthermore, the concept of $\alpha$-top dominance is akin to the notions listed above in the sense that for a semi-standard zero-normalised game $v \in \mathbb V^N_0$, property (\ref{eq:TopDom}) can be rewritten as
\[
\frac{v(S)}{\sum_{j \in S} M_j (v)^\alpha} \leqslant \frac{v(N)}{\sum_{j \in N} M_j (v)^\alpha}
\]
for $\sum_{j \in N} M_j (v)^\alpha \geqslant \sum_{j \in S} M_j (v)^\alpha > 0$. Moreover, implementing $\alpha =0$, the notion of $\alpha$-top dominance clearly generalises the notion of top convexity, as top convexity is equivalent to 0-top dominance for zero-normalised games. Indeed, for zero-normalised game $v \in \mathbb V^N_0$ we straightforwardly derive $\sum_{j \in S} M^0_j (v) = |S|$, immediately leading to the conclusion that 0-top dominance is the same as top convexity.

The next theorem generalises the insights of Theorem \ref{thm:3playerCore} to games with arbitrary player sets.
 
\begin{theorem} \label{thm:MainCore}
	Let $\alpha > 0$. A standard cooperative game $v \in \mathbb V^N$ is $\alpha$-top dominant if and only if $g^\alpha (v) \in C(v)$.
\end{theorem}
\begin{proof}
	Let $v \in \mathbb V^N$ be standard and let $\alpha > 0$. \\ Now $g^\alpha (v) \in C(v)$ if and only if it holds that for every coalition $S \in 2^N \colon \sum_{j \in S} g^\alpha_j (v) \geqslant v(S)$. This is equivalent to the condition that for every coalition $S \in 2^N \colon$
	\[
	\sum_{i \in S} v_i + \frac{\sum_{i \in S} \left( M_i (v) - v_i \, \right)^\alpha}{\sum_{j \in N} \left( M_j (v) - v_j \, \right)^\alpha} \cdot \left( v(N) - \sum_{j \in N} v_j \right) \geqslant v(S)
	\]
	From $v$ being standard, it follows that $\sum_{i \in N} \left( M_i (v) - v_i \, \right)^\alpha >0$. Hence, the above is equivalent to the condition that for every coalition $S \in 2^N \colon$
	\[
	\sum_{i \in S} \left( M_i (v) - v_i \, \right)^\alpha \cdot \left( v(N) - \sum_{j \in N} v_j \right) \geqslant \sum_{j \in N} \left( M_j (v) - v_j \, \right)^\alpha \cdot \left( v(S) - \sum_{i \in S} v_i \right)
	\]
	This is exactly the $\alpha$-top dominance property.
\end{proof}

\paragraph{Properties of top dominant games}

The next definition introduces a reduced notion of superadditivity that fits with top dominance. This form of superadditivity is defined as ``partitional'' superadditivity. 
\begin{definition}
	A cooperative game $v \in \mathbb V^N$ is \textbf{partitionally superadditive} if for every coalition $S \subseteq N$ it holds that $v(S) + v(N \setminus S) \leqslant v(N)$.
\end{definition} 
The next theorem shows that top dominant games always satisfy regularity as well as the partitional superadditivity property defined above.

\begin{theorem} \label{thm:Zeroregular}
Let $v \in \mathbb V^N$ be a standard cooperative game. 
If the game $v$ is $\alpha$-top dominant for some $\alpha > 0$, then $v$ is regular as well as partitionally superadditive.
\end{theorem}
\begin{proof}
	Let $v \in \mathbb V^N$ be a standard game and let $\alpha > 0$ be such that $v$ is $\alpha$-top dominant. Hence, $\sum_{j \in N} \left( M_v(j) - v_j \, \right)^\alpha >0$ and $\sum_{j \in S} \left( M_v(j) - v_j \, \right)^\alpha \geqslant 0$ for every coalition $S \subset N$.
	
	\bigskip\noindent
	We first show that $v$ is essential, together with the hypothesis that $v$ is standard, implying that $v$ is regular. \\
	When we apply the $\alpha$-top dominance property to the coalition $N-i$ for any $i \in N$ we arrive at
	\[
	\left( \, v(N-i) - \sum_{j \neq i } v_j \, \right) \cdot \sum_{j \in N} \left( M_v (j) -v_j \, \right)^\alpha \leqslant \left( \, v(N) - \sum_{j \in N } v_j \, \right) \cdot \sum_{j \neq i} \left( M_v(j) - v_j \, \right)^\alpha
	\]
	Adding these inequalities over all $i \in N$ we arrive at the conclusion that
	\begin{align*}
		\sum_{i \in N} \left( \, v(N-i) - \sum_{j \neq i } v_j \, \right) \cdot \sum_{j \in N} \left( M_v (j) -v_j \, \right)^\alpha & \leqslant \left( \, v(N) - \sum_{j \in N } v_j \, \right) \sum_{i \in N} \sum_{j \neq i} \left( M_v (j) -v_j \, \right)^\alpha \\
		& = (n-1) \left( \, v(N) - \sum_{j \in N } v_j \, \right) \sum_{j \in N} \left( M_v (j) -v_j \, \right)^\alpha
	\end{align*}
	Hence,
	\[
	\sum_{i \in N} \left( \, v(N-i) - \sum_{j \neq i } v_j \, \right) \leqslant (n-1) \left( \, v(N) - \sum_{j \in N } v_j \, \right)
	\]
	leading to the conclusion that
	\[
	\sum_{i \in N} [ v(N) - v(N-i) \, ] - v(N) - (n-1) \sum_{i \in N} v_i \geqslant - (n-1) \sum_{i \in N} v_i .
	\]
	This implies that
	\begin{equation} \label{eq:HalfEssential}
		\sum_{i \in N} M_i (v) \geqslant v(N) .
	\end{equation}
	Next, suppose to the contrary that $v(N) < \sum_{j \in N} v_j$.  From $v$ being a standard game, there is some $i \in N$ with $M_i (v) > v_i$. We can apply the $\alpha$-top dominance property to $S = \{ i \}$ and derive that
	\[
		0 = (v_i - v_i) \cdot \sum_{j \in N} ( M_j(v) - v_j )^\alpha \leqslant \left( v(N) - \sum_{j \in N} v_j \right) \cdot (M_i(v) - v_i) < 0
	\]
	which is impossible. Therefore, we conclude that $v(N) \geqslant \sum_{j \in N} v_j$ and, together with (\ref{eq:HalfEssential}), we have shown the assertion that $v$ is essential.
	
	\bigskip\noindent
	Next we show that $v$ is partitionally superadditive. \\ Let $S \subset N$ be some coalition. Then, from $\alpha$-top dominance, it holds for $S$ that
	\begin{align*}
		\left( v(S) - \sum_{j \in S} v_j \right) \cdot \sum_{i \in N} \left( M_i (v) - v_i \, \right)^\alpha & \leqslant \left( v(N) - \sum_{j \in N} v_j \right) \cdot \sum_{i \in S} \left( M_i (v) - v_i \, \right)^\alpha \\
		\left( v(N \setminus S) - \sum_{j \in N \setminus S} v_j \right) \cdot \sum_{i \in N} \left( M_i (v) - v_i \, \right)^\alpha & \leqslant \left( v(N) - \sum_{j \in N} v_j \right) \cdot \sum_{i \in N \setminus S} \left( M_i (v) - v_i \, \right)^\alpha
	\end{align*}
	Adding these two inequalities leads to the conclusion that
	\[
	\left( v(S) + v(N \setminus S) - \sum_{j \in N} v_j \right) \cdot \sum_{i \in N} \left( M_i (v) - v_i \, \right)^\alpha \leqslant \left( v(N) - \sum_{j \in N} v_j \right) \cdot \sum_{i \in N} \left( M_i (v) - v_i \, \right)^\alpha
	\]
	Since $\sum_{i \in N} \left( M_i (v) - v_i \, \right)^\alpha >0$ for any $\alpha > 0$, we have shown that 
	\[
	v(S) + v(N \setminus S) - \sum_{j \in N} v_j \leqslant v(N) - \sum_{j \in N} v_j
	\]
	and, hence, $v(S) + v(N \setminus S) \leqslant v(N)$. We conclude that $v$ is indeed partitionally superadditive. 
\end{proof}

\bigskip\noindent
Theorems \ref{thm:Zeroregular} and \ref{thm:MainCore} now immediately imply the following corollary.
\begin{corollary} \label{coro:CoreSuperadditive}
	Let $v \in \mathbb V^N$ be a standard cooperative game and let $\alpha > 0$. If $g^\alpha (v) \in C(v)$, then $v$ is regular and partitionally superadditive. 
\end{corollary}
One can ask oneself whether the condition of top dominance can be simplified or linked to other regularity properties of cooperative games. As shown in Theorem \ref{thm:Zeroregular} it is clear that top dominance is closely related to the superadditivity property that is widely used in cooperative game theory. The next example shows that top dominance is actually strictly weaker than superadditivity.
\begin{example}
	Consider a regular and zero-normalised three-player game with $N = \{ 1,2,3 \}$ and $v$ given by $v_i =0$ for $i = 1,2,3$, $v(12) = v(13) =-1$, $v(23)=0$ and $v(N)=1$. We note that $v$ is not superadditive, since $v_1+v_2=0 > v(12)=-1$. 
	\\
	However, for any $\alpha >0$ we remark that
	\begin{align*}
		\frac{v(N)}{M_1(v)^\alpha + M_2(v)^\alpha + M_3 (v)^\alpha} & = \frac{1}{1+ 2^{\alpha +1}} >0 \\
		\frac{v(12)}{M_1(v)^\alpha + M_2(v)^\alpha} & = \frac{-1}{1+ 2^{\alpha}} <0 \\
		\frac{v(13)}{M_1(v)^\alpha + M_3 (v)^\alpha} & = \frac{-1}{1+ 2^{\alpha}} <0 \\
		\frac{v(23)}{M_2(v)^\alpha + M_3 (v)^\alpha} & = 0
	\end{align*}
	Hence, we conclude that $v$ indeed satisfies $\alpha$-top dominance for every $\alpha > 0$.
	\\
	Furthermore, we determine easily that $M_1 (v) =1$ and $M_2 (v) = M_3 (v) =2$, leading to the conclusion that for every $\alpha >0$ the Gately values are given as $g^\alpha_1 = \frac{1}{1+ 2^{\alpha +1}}$ and $g^\alpha_2 = g^\alpha_3 = \frac{2^\alpha}{1+ 2^{\alpha +1}}$.
	\\
	It can also easily be checked that for every $\alpha >0 \colon g^\alpha (v) \in C(v)$. 
\end{example}

\subsection{Comparing the Gately and Shapley values}

The Shapley value \citep{Shapley1953} has achieved the status as being the prime solution concept for TU-games. It sets a benchmark for assessing the suitability of alternative solution concepts. In this context it is suitable to consider for what classes of TU-games such alternative solution concepts lead to exactly the same allocation as the Shapley value. We pursue that here for the Gately solution concept.

We show that the Shapley and Gately values coincide on narrow, but highly relevant, classes of games that satisfy rather strong regularity properties. Theorem \ref{thm:Gately=Shapley} below identifies a class of structured games that have rather wide applicability. This class is founded on strong conditions on the class of constituting coalitions and the corresponding unanimity games

We recall for the benefit of the next analysis that all cooperative games can be represented through the \emph{unanimity basis} of $\mathbb V^N$. For every coalition $S \in 2^N$ we define the $S$-\emph{unanimity game} $u_S \in \mathbb V^N$ by
\[
u_S (T) = \left\{
\begin{array}{ll}
	1 & \mbox{if } S \subseteq T \\
	0 & \mbox{otherwise}
\end{array}
\right.
\]
Now, every game $v \in \mathbb V^N$ can be written as $v = \sum_{S \in \Pi_v} \Delta_S (v) \cdot u_S$, where $\Pi_v = \{ S \in 2^N \mid \Delta_S (v) \neq 0 \}$ is the class of relevant constituting coalitions with $u_S \in \mathbb V^N$ being the corresponding $S$-unanimity game for the constituting coalition $S \in \Pi_v$ and $\Delta_v (S) \neq 0$ the corresponding Harsanyi dividend of coalition $S$ in game $v$ \citep{Harsanyi1959}.

We also recall that for any game $v \in \mathbb V^N$ with corresponding representation $v = \sum_{S \in \Pi_v} \Delta_S (v) \cdot u_S$, the \emph{Shapley value} of $v$ is defined as $\phi (v) \in \mathbb R^N$ given by
\begin{equation}
	\phi_i (v) = \sum_{S \in 2^N \colon i \in S} \, \frac{\Delta_v (S)}{|S|} \qquad \mbox{for every } i \in N.
\end{equation}
The next definition introduces some relevant classes of regular games.

\begin{definition}
	Let $N = \{ 1, \ldots ,n \}$ be a set of players and let $k \in \{ 2, \ldots , n-1 \}$.
	A game $v \in \mathbb V^N$ on $N$ is denoted as a $\mathbf k$-\textbf{game} if $v$ is regular and $v$ can be written as $v = \sum_{S \in \Pi_v} \Delta_v (S) \, u_S$, where $\Pi_v = \{ S \in 2^N \mid \Delta_S (v) \neq 0 \}$, such that $| S | =k$ for all constituting coalitions $S \in \Pi_v$. \\ The subclass of $k$-games on $N$ is denoted by $\mathbb V^N_k \subset \mathbb V^N$.
\end{definition}
The subclass of 2-games has been investigated in the literature on its properties. In particular, \citet{Nouweland1996} and \citet{Brink2023} show that for 2-games the Shapley value coincides with the Nucleolus and the $\tau$-value. It might not be a surprise that we can show for all subclasses of $k$-games the property that the Gately value coincides with the Shapley value.
\begin{theorem} \label{thm:Gately=Shapley}
	Let $N = \{ 1, \ldots ,n \}$ be a set of players and let $k \in \{ 2, \ldots , n-1 \}$. For every $k$-game $v \in \mathbb V^N_k$ it holds that $g(v) = \phi (v)$.
\end{theorem}

\begin{proof}
	Let $N = \{ 1, \ldots ,n \}$ be a set of players and let $k \in \{ 2, \ldots , n-1 \}$. First, we remark that all $k$-games are zero-normalised by definition. Now, let $v \in \mathbb V^N_k$ be a $k$-game. Hence, $|S|=k$ for all $S \in \Pi_v$. Assume that $| \Pi_v | =m$ is the number of the constituting coalitions of the game $v$.  \\ Next, we introduce some additional notation. For every $i \in N$ we let $\Pi_i = \{ S \in \Pi_v \mid i \in S \}$. Then the Shapley value of player $i \in N$ can be written as
	\[
	\phi_i (v) = \sum_{S \in \Pi_i} \frac{\Delta_v (S)}{|S|} = \tfrac{1}{k} \sum_{S \in \Pi_i} \Delta_v (S) = \frac{\Delta_i}{k} \qquad \mbox{where } \Delta_i = \sum_{S \in \Pi_i} \Delta_v (S) .
	\]
	Regarding the determination of the Gately value, we note that the marginal contribution of $i \in N$ is now given by
	\[
	M_i (v) = v(N) - v(N-i) = \sum_{S \in \Pi_v} \Delta_v (S) - \sum_{S \in \Pi_v \colon i \notin S} \Delta_v (S) = \sum_{S \in \Pi_i} \Delta_v (S) = \Delta_i .
	\]
	Furthermore, this implies that $\sum_{j \in N} M_i (v) = \sum_{j \in N} \Delta_j = k \cdot v(N)$. Hence, we have determined that for $i \in N \colon$
	\[
	g_i (v) = \frac{M_i (v)}{\sum_{j \in N} M_j (v)} \cdot v(N) = \frac{\Delta_i}{k \cdot v(N)} \cdot v(N) = \frac{\Delta_i}{k} = \phi_i (v) .
	\]
	This shows the assertion of the theorem.
\end{proof}

\bigskip\noindent
Based on Theorem 1 and Corollary 1 of \citet{Brink2023} in combination with Theorem \ref{thm:Gately=Shapley}, we can determine another characterisation of the Gately value for the subclass of 2-games:
\begin{corollary}
	On the subclass of 2-games $\mathbb V^N_2$, the Gately value is the unique value that satisfies the balanced externalities property in the sense that for every $v \in \mathbb V^N_2$ with $v = \sum_{S \in \Pi_v} \Delta_v (S) \, u_S \colon$
	\begin{equation}
		g_i (v) = \sum_{j \neq i} \left( \, g_j (v) - g_j ( v^{-i} ) \, \right)
	\end{equation}
	where $v^{-i} = \sum_{T \in \Pi^{-i}_v} \Delta_v (T) \, u_T$ where $\Pi^{-i}_v = \{ T \in \Pi_v \mid i \notin T \}$.
\end{corollary}
The reverse of Theorem \ref{thm:Gately=Shapley} does not hold. There are other classes of games with strong regularity properties that are not $k$-games and on which the Gately value coincides with the Shapley value. The next proposition introduces such a subclass of highly regular games.
\begin{proposition} \label{prop:Gately=Shapley}
	Let $v \in \mathbb V^N$ be a regular game written as $v = \sum_{S \in \Pi_v} \Delta_v (S) \, u_S$. Assume that $n = k \cdot m$, where $k,m \in \mathbb N$ with $k \neq m$, such that $\Pi_v = \Pi^k \biguplus \Pi^m$, where $\Pi^k$ is a partitioning of $N$ into $m$ sets of size $k$ and $\Pi^m$ is a partitioning of $N$ into $k$ sets of size $m$. \\ If there exists some $\Delta \neq 0$ such that $\Delta_v (S) = \Delta$ for all $S \in \Pi_v$, then $g(v) = \phi (v)$.
\end{proposition}

\begin{proof}
	Let $i \in N$ be an arbitrary player. \\ Due to the structure of the game, player $i \in N$ is member of exactly one coalition $S \in \Pi^k$ and one coalition $T \in \Pi^m$. Hence, it easily follows that $\phi_i (v) = \tfrac{\Delta}{k} + \tfrac{\Delta}{m}$. \\ To compute the Gately value, we note that $M_i (v) =2 \Delta$ and $\sum_{j \in N} M_j(v) = n \cdot 2 \Delta = 2km \Delta$. Furthermore, $v(N) = (k+m) \Delta$, so the Gately value for player $i \in N$ can be computed as
	\[
	g_i (v) = \frac{2 \Delta}{2km \Delta} \, (k+m) \Delta =\frac{k+m}{km} \Delta = \tfrac{\Delta}{k} + \tfrac{\Delta}{m} = \Phi_i (v) .
	\]
	This shows the assertion.
\end{proof}

\bigskip\noindent
For games that do not satisfy the property stated in Theorem \ref{thm:Gately=Shapley}, the Gately value is only very rarely equal to the Shapley value. The next example discusses a convex five-player game in which the Gately value is not in the Core, while the Shapley value is a Core selector \citep{Shapley1971}.

\begin{example}
	Let $N = \{ 1,2,3,4,5 \}$ and let $v = u_{12} + 3 u_{345}$. This game is convex with $v(N) = 4$ and the marginal contribution vector $M = (1,1,3,3,3)$. It is easy to see that the Shapley value is given by $\phi = \left( \, \tfrac{1}{2} \, , \, \tfrac{1}{2} \, , \, 1,1,1 \, \right)$ and the Gately value is given by $g = \left( \tfrac{4}{11} \, , \, \tfrac{4}{11} \, , \, \tfrac{12}{11} \, , \, \tfrac{12}{11} \, , \, \tfrac{12}{11} \,  \right)$. \\ We note that $\phi \in C(v)$, while $g \notin C(v)$ since $g_1 + g_2 = \tfrac{8}{11} < 1 = v(12)$. \\ We remark that the game $v$ discussed here does not satisfy either of the descriptions introduced in Theorem \ref{thm:Gately=Shapley} and Proposition \ref{prop:Gately=Shapley}, but that the game has a structure that is similar to the structure of the class of games considered in Proposition \ref{prop:Gately=Shapley}.
\end{example}

\newpage
\singlespace
\bibliographystyle{ecta}
\bibliography{RPDB}

\end{document}